\documentclass[acmsmall,screen]{acmart}

\AtBeginDocument{%
  \providecommand\BibTeX{{%
    \normalfont B\kern-0.5em{\scshape i\kern-0.25em b}\kern-0.8em\TeX}}}

\usepackage{adjustbox}
\usepackage{amsmath,amssymb,amsfonts}
\usepackage{algorithmic}
\usepackage{graphicx}
\usepackage{textcomp}
\usepackage{xcolor}
\usepackage{multirow}
\usepackage{mdframed}
\usepackage{tcolorbox}
\usepackage{makecell}
\usepackage{threeparttable}
\usepackage{colortbl}
\usepackage{subfigure}
\usepackage{hhline}
\usepackage{hyperref}
\usepackage{pifont}
\usepackage{enumitem}
\usepackage{caption}

\usepackage{bm}
\usepackage{balance}

\newcommand{\parabf}[1]{\noindent\textbf{#1}}

\newcommand{\Comment}[1]{}
\newcommand{\yiling}[1]{\textcolor{blue}{#1}}

\newcommand{\lingming}[1]{\textcolor{red}{#1}}
\newcommand{\revised}[1]{\textcolor{black}{#1}}

\newcommand{\apr}{automated program repair}

\definecolor{ggray}{HTML}{eff0f0}
\definecolor{gggray}{HTML}{E8E8E8}
\definecolor{ggggray}{HTML}{BEBEBE}

\newcommand{\prog}{\mathcal{P}_b}
\newcommand{\patch}{\mathcal{P}}
\newcommand{\patches}{\mathbb{P}}
\newcommand{\tst}{t}

\newcommand{\rtsMatrix}{\mathbb{M}^{RTS}}

\newcommand{\Nreduction}{Reduction}

\newcommand{\testnum}[2]{NT(#1)}

\newcommand{\pMatrixpart}{\mathbb{M}_p}
\newcommand{\pMatrixfull}{\mathbb{M}_f}
\newcommand{\tsts}{\mathcal{T}}
\newcommand{\tstsrts}{\mathcal{T'}}
\newcommand{\unknown}{\texttwelveudash}%
\newcommand{\skips}{\ding{109} } %
\newcommand{\pass}{\ding{51}}%
\newcommand{\fail}{\ding{55}}%
\newcommand{\pMatrix}{\mathbb{M}}
\newcommand{\pMatrixCell}[2]{\mathbb{M}[#1,#2]}

\newcommand{\modcode}{\patch_{\Delta}}
\newcommand{\covcode}{\mathbb{C}[\prog,\tst]}

\newcommand{\etal}{\textit{et al.}}

\newcommand{\rts}{RTS}
\newcommand{\rtsclass}{$RTS_{class}$}
\newcommand{\rtsmethod}{$RTS_{method}$}
\newcommand{\rtsstate}{$RTS_{stmt}$}
\newcommand{\norts}{$RTS_{no}$}

\newcommand{\priotrp}{$TP_{APR}$}
\newcommand{\priodefault}{$TP_{base}$}

\newcommand{\priototal}{$RTP_{tot}$}
\newcommand{\prioadd}{$RTP_{add}$}

\newcommand{\stopfpatch}{$\patches_{F2F}$}
\newcommand{\stopppatch}{$\patches_{P2F}$}
\newcommand{\plausiblepatch}{$\patches_{\checkmark}$}

\newcommand{\mcpatch}{$\patches_{MC}$}
\newcommand{\scpatch}{$\patches_{SC}$}
\newcommand{\mmpatch}{$\patches_{MM}$}
\newcommand{\smpatch}{$\patches_{SM}$}
\newcommand{\mspatch}{$\patches_{MS}$}
\newcommand{\sspatch}{$\patches_{SS}$}

\newcommand{\nti}{$NT_{i}$}
\newcommand{\ntacc}{$NT_{acc}$}
\newcommand{\ntave}{$NT_{ave}$}

\newcommand{\acs}{ACS}
\newcommand{\prapr}{PraPR}

\newcommand{\arja}{Arja}
\newcommand{\kali}{Kali-A}
\newcommand{\genprog}{GenProg-A}
\newcommand{\rsrepair}{RSRepair-A}

\newcommand{\cardumen}{Cardumen}
\newcommand{\jgenprog}{jGenProg}
\newcommand{\jKali}{jKali}
\newcommand{\jmutrepair}{jMutRepair}

\newcommand{\dynamoth}{Dynamoth}

\newcommand{\avatar}{AVATAR}
\newcommand{\kpar}{kPar}
\newcommand{\fixminer}{FixMiner}
\newcommand{\tbar}{TBar}

\newcommand{\simfix}{SimFix}
\newcommand{\capgen}{CapGen}

\newcommand{\trp}{TrpAutoRepair}
\newcommand{\ourtool}{OURTOOL}

\usepackage[utf8]{inputenc}
\usepackage[english]{babel}

\newtheorem{theorem}{\textbf{Definition}}[section]

\newcounter{finding}
\newcommand{\finding}[1]{\refstepcounter{finding}
 	\vspace{5mm}
	\begin{mdframed}[linecolor=gray,roundcorner=12pt,backgroundcolor=gray!15,linewidth=3pt,innerleftmargin=2pt, leftmargin=0cm,rightmargin=0cm,topline=false,bottomline=false,rightline = false]
		\textbf{Finding \arabic{finding}:} #1
	\end{mdframed}
	\vspace{5mm}
}

\newcommand{\distance}{5pt}
\begin{document}

\title{When Automated Program Repair Meets Regression Testing  \\ -- An Extensive Study on 2 Million Patches}


\author{Yiling Lou}
\email{yilinglou@fudan.edu.cn}
\affiliation{%
  \institution{Fudan University}
  \city{Shanghai}
  \country{China}}

\author{Jun Yang}
\affiliation{%
  \institution{University of Illinois Urbana-Champaign}
  \city{Champaign}
  \country{USA}
}

\author{Samuel Benton}
\affiliation{%
  \institution{The University of Texas at Dallas}
  \city{Dallas}
  \country{USA}
}

\author{Dan Hao}
\affiliation{%
  \institution{Peking University}
  \city{Beijing}
  \country{China}}

\author{Lin Tan}
\affiliation{%
  \institution{Purdue University}
  \city{West Lafayette}
  \country{USA}}

\author{Zhenpeng Chen}
\affiliation{%
  \institution{Nanyang Technological University}
  \country{Singapore}}

\author{Lu Zhang}
\affiliation{%
  \institution{Peking University}
  \city{Beijing}
  \country{China}}

\author{Lingming Zhang}
\affiliation{%
  \institution{University of Illinois Urbana-Champaign}
  \city{Champaign}
  \country{USA}
}

\renewcommand{\shortauthors}{Lou et al.}

\begin{abstract}
In recent years, Automated Program Repair (APR) has been extensively studied in academia and even drawn wide attention from industry. However, APR techniques can be extremely time consuming since (1) a large number of patches can be generated for a given bug, and (2) each patch needs to be executed on the original tests to ensure its correctness. In the literature, various techniques (e.g., based on learning, mining,  and constraint solving) have been proposed/studied to reduce the number of patches. Intuitively, every patch can be treated as a software revision during regression testing; thus, traditional Regression Test Selection (RTS) techniques can be leveraged to only execute the tests affected by each patch (as the other tests would keep the same outcomes) to further reduce patch execution time. However, few APR systems actually adopt RTS and there is still a lack of systematic studies demonstrating the benefits of RTS and the impact of different RTS strategies on APR. To this end, this paper presents the first extensive study of widely-used RTS techniques at different levels (i.e., class/method/statement levels) for 12 state-of-the-art APR systems on over 2M patches. Our study reveals various practical guidelines for bridging the gap between APR and regression testing, including: (1) the number of patches widely used for measuring APR efficiency can incur skewed conclusions, and the use of inconsistent RTS configurations can further skew the conclusions; (2) all studied RTS techniques can substantially improve APR efficiency and should be considered in future APR work; (3) method- and statement-level RTS outperform class-level RTS substantially, and should be preferred; (4) RTS techniques can substantially outperform state-of-the-art test prioritization techniques for APR, and combining them can further improve APR efficiency; and (5) traditional Regression Test Prioritization (RTP)  widely studied in regression testing performs even better than APR-specific test prioritization when combined with most RTS techniques. Furthermore, we also present the detailed impact of different patch categories and patch validation strategies on our findings.
\end{abstract}



\authorsaddresses{Authors’ addresses: 
Yiling Lou is with the Department of Computer Science, Fudan University, 2005 Songhu Road, Shanghai 200437, China. E-mail: yilinglou@fudan.edu.cn.
Jun Yang and Lingming Zhang are with the Department of Computer Science, University of Illinois Urbana-Champaign, 506 S. Wright St. Urbana, IL 61801-3633, USA. E-mails: \{jy70@, lingming\}@illinois.edu.
Samuel Benton is with the Department of Computer Science, The University of Texas at Dallas, 800 W Campbell Rd, Richardson, TX 75080, USA. E-mail: samuel.benton1@utdallas.edu.
Dan Hao and Lu Zhang are with the School of Computer Science, Peking University, No. 5 Yiheyuan Road, Beijing 100871, China. E-mails: \{haodan, zhanglucs\}@pku.edu.cn. 
Lin Tan is with the Department of Computer Science, Purdue University, 610 Purdue Mall, West Lafayette, IN 47907, USA. E-mail: lintan@purdue.edu.
Zhenpeng Chen (Corresponding author) is with the School of Computer Science and Engineering, Nanyang Technological University, 50 Nanyang Ave, Singapore 639798, Singapore. E-mail: zhenpeng.chen@ntu.edu.sg.}


\maketitle

\section{Introduction} \label{sec:intro}
\Comment{Software bugs can trigger unexpected program behaviors and result in fatal consequences. e.g., causing the  financial loss or threatening the personal safety. Manual debugging is often tedious and costly.}

\textbf{Automated Program Repair (APR)} ~\cite{kechagia2021evaluating,DBLP:conf/icse/XiongWYZH0017/acs, DBLP:conf/icse/WenCWHC18, DBLP:journals/tse/YuanB20/arja,DBLP:journals/ese/KoyuncuLBKKMT20/fixminer, DBLP:conf/icse/JiangLLTGZ23} 
has been proposed to automatically generate patches for buggy programs so as to reduce manual debugging efforts.
Modern APR techniques often follow a \emph{generate-and-validate} procedure, leveraging
test suites as the partial specification of the desired program behavior.
More specifically, a test-based APR system repeatedly generates patches and validates them against the whole test suite until a patch that can pass all tests is found.
To date, a large number of APR techniques have been proposed, effectively fixing a considerable number of real bugs and improving software quality/productivity~\cite{ DBLP:conf/issta/GhanbariBZ19,DBLP:conf/issta/JiangXZGC18}.

Although receiving wide attention from both academia and industry, APR techniques can be extremely time consuming~\cite{DBLP:conf/icse/HuaZWK18/sketchfix, DBLP:journals/sqj/GouesFW13, DBLP:conf/icse/LongR16,DBLP:conf/kbse/WeimerFF13, DBLP:conf/issta/GhanbariBZ19,DBLP:conf/icse/0001WKKB0WKMT20}: (1) for a given bug, there are a vast number of patches generated; (2) for each generated patch, it often takes non-trivial time to execute the original tests for correctness validation. \revised{Efficiency is essential for the practical usage of APR systems~\cite{DBLP:conf/icse/0001WKKB0WKMT20}, as both the production cycle and the development cycle require low-latency debugging assistance from APR systems. In particular, given the computing resources are not always sufficient, parallel execution cannot always be available for APR systems.}

To reduce overheads in the repair procedure,  researchers have proposed various cost reduction approaches. 
For example, numerous APR techniques have been proposed to reduce the number of generated patches, including constraint-based~\cite{DBLP:conf/icse/XiongWYZH0017/acs, DBLP:conf/ssbse/MartinezM18/cardumen, DBLP:conf/icse/DurieuxM16/dynamoth}, heuristic-based~\cite{DBLP:journals/tse/YuanB20/arja, DBLP:conf/issta/JiangXZGC18, DBLP:conf/icse/Wen0C20}, template-based~\cite{DBLP:conf/wcre/LiuK0B19/avatar, DBLP:journals/ese/KoyuncuLBKKMT20/fixminer,DBLP:conf/icst/LiuKB0KT19/kpar}, and learning-based techniques~\cite{DBLP:conf/issta/LutellierPPLW020/coconut, chen2019sequencer, DBLP:conf/icse/Li0N20/dlfix}. However, APR still remains one of the most costly approaches in software engineering. \Comment{
\Comment{For example, Chen~\etal{}~\cite{DBLP:conf/kbse/Chen0F17/jaid} and Hua~\etal{}~\cite{DBLP:conf/icse/HuaZWK18/sketchfix} accelerate each test execution by abstracting program with meta-states and sketches. Ghanbari~\etal{}\cite{DBLP:conf/issta/GhanbariBZ19} generate patches directly at byte-code level so as to save patch compilation and loading time.}
Orthogonal to these optimizations, another possible solution  is to optimize test executions within each patch validation, e.g., reducing the number of test executions in  each patch validation. }

\parabf{Regression Testing}~\cite{DBLP:journals/stvr/YooH12} reruns regression tests on every software revision to check whether it breaks previously working functionalities, and has been widely adopted in practice. Meanwhile, rerunning all tests for each revision can be extremely time consuming~\cite{DBLP:conf/sigsoft/ElbaumRP14, DBLP:conf/icse/MemonGNDNSM17}. Therefore, researchers have proposed various regression testing techniques to speed up the process. For example, given a software revision, Regression Test Selection (RTS)~\cite{DBLP:conf/sigsoft/LegunsenHSLZM16,DBLP:conf/sigsoft/OrsoSH04, DBLP:conf/issta/GligoricEM15,DBLP:journals/tosem/RothermelH97,DBLP:conf/icsm/RothermelH93} only selects/executes the tests affected by code changes for faster regression testing (because the other tests should have the same outcomes on the original and new revisions).
To date, RTS techniques have been shown to significantly accelerate regression testing and has been widely incorporated into build systems of open-source projects~\cite{DBLP:conf/issta/GligoricEM15, DBLP:conf/icse/Zhang18} and commercial systems~\cite{DBLP:conf/sigsoft/CelikLG18, DBLP:conf/icst/ZhongZK19}. 
\Comment{such as regression test selection (RTS)~\cite{DBLP:conf/sigsoft/LegunsenHSLZM16,DBLP:conf/sigsoft/OrsoSH04, DBLP:conf/issta/GligoricEM15,DBLP:journals/tosem/RothermelH97,DBLP:conf/icsm/RothermelH93}, regression test prioritization (RTP)~\cite{DBLP:journals/tse/LiHH07, DBLP:journals/ac/Lou0ZH19, DBLP:conf/icsm/QiML13}, and test suite reduction (TSR)~\cite{}. RTS only selects/executes the tests affected by code changes for faster regression testing (because the other tests should have the same outcomes on the original and new revisions), while RTP and TSR techniques aims to prioritize and reduce the existing tests for faster bug detection. Among the three categories of regression techniques, RTS has been shown to significantly accelerate regression testing and has been widely incorporated into build systems of open-source projects~\cite{DBLP:conf/issta/GligoricEM15, DBLP:conf/icse/Zhang18}.}

In fact, \emph{APR shares a similar procedure with regression testing that the buggy program is repeatedly modified (by APR systems rather than developers during regression testing) and each generated patch can be treated as a software revision to be exhaustively validated during regression testing.} Therefore, regression testing techniques, such as RTS, can be naturally applied to accelerate APR. However, surprisingly,
after systematically revisiting the APR literature, we find that such an important optimization receives little attention from existing APR work: \textit{most existing APR systems do not apply any RTS, while the few APR systems doing so adopt RTS at different granularities without demonstrating their clear benefits}. For example, \prapr{}~\cite{DBLP:conf/issta/GhanbariBZ19} adopts statement-level RTS while \capgen{}~\cite{DBLP:conf/icse/WenCWHC18} uses class-level RTS in its implementation without any justification.
Therefore, \textit{it remains unknown that how important it is to adopt RTS in APR systems and how different RTS techniques affect APR efficiency}.
As a result, 16 of the 17 APR systems proposed in recent three years have still not adopted any RTS at all (Section~\ref{sec:revisit}).

To bridge such a knowledge gap, we perform the first extensive study to investigate the impact of widely-used RTS techniques on APR efficiency. \Comment{In this work, we mainly focus on RTS, an important category of regression testing techniques.} More specifically, we first investigate the cost reduction achieved by widely-used RTS techniques at different levels (i.e., class/method/statement levels); furthermore, we also study the joint impact of RTS and other popular regression testing techniques (i.e., various test prioritization techniques which reorder test executions and can also potentially speed up patch validation) on APR efficiency.
Our study is conducted on 12 state-of-the-art APR systems with over 2M patches generated for the widely-used Defects4J~\cite{DBLP:conf/issta/JustJE14} benchmark.

Our study reveals various practical guidelines for the APR community. 
(1) The number of patches widely used for measuring APR efficiency can incur skewed conclusions, and the use of inconsistent RTS configurations can further skew the conclusion.
\Comment{The widely-used metric (i.e., the number of generated patches) and the inconsistent RTS configurations can both induce biased conclusions on measuring APR efficiency.}
(2) RTS can indeed reduce a significant portion of test executions on all studied APR systems and should be considered in future APR work. 
(3) The performance varies among different RTS strategies. In particular, statement- and method-level RTS can significantly outperform class-level RTS, and are recommended for future APR work. 
(4) RTS techniques can substantially outperform state-of-the-art test prioritization techniques for APR, and combining them can further improve APR efficiency.
(5) Traditional Regression Test Prioritization (RTP) techniques~\cite{DBLP:journals/tse/LiHH07, DBLP:journals/ac/Lou0ZH19} widely studied in regression testing performs even better than APR-specific test prioritization~\cite{DBLP:conf/icsm/QiML13} when they are combined with most RTS techniques.
\Comment{Combined with class-level or method-level RTS in APR, traditional regression test prioritization widely studied in regression testing outperforms state-of-the-art test prioritization specifically designed for APR.}
Besides, we further present the detailed impact of different patch categories and patch validation strategies on our findings. Lastly, we also discuss the impact of RTS strategies on the repair effectiveness. 

\Comment{, and observed a more remarkable impact of RTS on the patches of stronger fixing capabilities and smaller fixing code scales}
As the first extensive study on the impact of regression testing on APR, this paper makes the following contributions:
\begin{itemize}
    \item \textbf{Literature review.} Revisiting the literature and implementations of existing APR systems, highlighting RTS as an important optimization mostly neglected and inconsistently-configured by existing work.
    
    
    \item \textbf{Extensive study.} Performing the first extensive study on the impact of different regression testing strategies on APR efficiency, exploring various representative RTS techniques (and RTP techniques) for 12 state-of-the-art APR systems on a well-established benchmark (Defects4J) with 395 real bugs, involving over 2M patches in total.

    
    \item \textbf{Practical guidelines.} Demonstrating the importance of  regression testing for APR and revealing various practical guidelines regarding the adoption of regression testing for the APR community and future APR work.\Comment{: RTS significantly improves APR efficiency and should be used by future work; RTS at finer granularities is preferred; future work should consider the number of test executions and align RTS configurations when measuring APR efficiency; combining RTS with test prioritization can future improve APR efficiency.}

\end{itemize}

\section{Background and Motivation}\label{sec:motivation}

\subsection{Regression Test Selection}
\Comment{
Regression testing~\cite{DBLP:journals/stvr/YooH12} is important in software evolution since it ensures the previously working functionality is not broken in the new revision. However, rerunning all tests for each new revision can be extremely time consuming~\cite{}. Therefore, researchers have proposed various optimization strategies to accelerate regression testing. Among them, regression test selection, regression test prioritization, and test suite reduction are the three most common techniques.}

Regression Test Selection (RTS)~\cite{DBLP:conf/sigsoft/LegunsenHSLZM16,DBLP:conf/sigsoft/OrsoSH04, DBLP:conf/issta/GligoricEM15,DBLP:journals/tosem/RothermelH97,DBLP:conf/icsm/RothermelH93}  accelerates regression testing by executing only a subset of regression tests. Its basic intuition is that the tests not affected by code changes would have the same results on the original and modified revisions.
A RTS technique is regarded as \textit{safe} if it selects all tests that may be affected by changed code, because missing any of such tests may fail to detect some regression bugs. 
\Comment{Test Suite Reduction (TSR)~\cite{} finds a subset of the original test suite according to certain testing requirements (e.g., coverage~\cite{}) and permanently removes the redundant test cases to reduce the testing costs. 
Note that TSR is not guaranteed to be safe and may reduce the fault-detection capability of the original test suite. Therefore, TSR is not included in this work since accuracy is essential for APR systems.
Regression Test Prioritization (RTP)~\cite{DBLP:conf/icse/HenardPHJT16,DBLP:conf/issta/EpitropakisYHB15} is proposed to detect regression faults faster by scheduling the execution order of existing regression test cases. By executing the tests that are more likely to fail first, RTP can reduce the testing costs and advance the time for developers to debug. Since in this work we mainly focus on the impact of RTS on APR, we then explain it in detail.}
A typical RTS strategy involves analyzing two dimensions of information: (1) \textit{changed code elements} between the original and modified program revisions, and (2) \textit{test dependencies} on the original revision. RTS then selects the tests whose dependencies overlap with the changed code elements. 

Based on the granularities of changed code elements and test dependencies, RTS techniques can be categorized as \textit{class-level}~\cite{DBLP:conf/issta/GligoricEM15,DBLP:conf/sigsoft/LegunsenHSLZM16}, \textit{method-level}~\cite{DBLP:conf/oopsla/RenSTRC04}, and \textit{statement-level} RTS~\cite{DBLP:conf/oopsla/HarroldJLLOPSSG01,DBLP:conf/sigsoft/OrsoSH04,DBLP:journals/tosem/RothermelH97}. For example, Gligoric~\etal{}~\cite{DBLP:conf/issta/GligoricEM15}  proposed an efficient class-level RTS, and  Zhang~\cite{DBLP:conf/icse/Zhang18} proposed a RTS strategy of hybrid granularities.
In addition, according to how test dependencies are analyzed, RTS techniques can be categorized as \textit{dynamic}~\cite{DBLP:conf/issta/GligoricEM15, DBLP:conf/oopsla/HarroldJLLOPSSG01,DBLP:conf/sigsoft/OrsoSH04} and \textit{static} RTS~\cite{DBLP:conf/sigsoft/LegunsenHSLZM16, DBLP:journals/joop/KungGHLT95}. 
For example, Legunsen \etal{}~\cite{DBLP:conf/sigsoft/LegunsenHSLZM16} compared the performance and safety of static regression test selection approaches in modern software evolution, and found that class-level static RTS was comparable to dynamic class-level RTS.
To date, RTS has been shown to substantially reduce the end-to-end regression testing costs~\cite{DBLP:conf/issta/GligoricEM15, DBLP:conf/icse/Zhang18} and has been widely incorporated into build systems of open-source projects~\cite{apachecamel,apachemath,apachecxf}.
While existing work has extensively studied different RTS approaches in the traditional regression testing and other related scenarios (e.g., non-functional genetic improvement~\cite{DBLP:conf/icse/GuizzoPSH21}), in this work, we perform the first extensive study of RTS in the APR scenario.

\subsection{Automated Program  Repair}~\label{sec:apr}
Automated Program Repair (APR)~\cite{DBLP:conf/ssbse/MartinezM18/cardumen, DBLP:conf/sigsoft/SmithBGB15, DBLP:conf/icse/YiTMBR18} automatically fixes program bugs with minimal human intervention. Given a buggy program, APR generates patches and then validates them to check their correctness. 
A typical test-based APR system takes a buggy program  and its test suite (with at least one failed test) as inputs, and consists of three phases.
(1) \textbf{Fault localization}: before the repair process, off-the-shelf fault localization ~\cite{abreu2007accuracy} is leveraged to diagnose suspicious code elements (e.g., statements).
(2) \textbf{Patch generation}: the APR system then applies repair operations on suspicious locations following the suspiciousness ranking list. Each modified program version is denoted as a \textit{candidate patch}.
(3) \textbf{Patch validation}: lastly, each candidate patch is validated by the test suite until a patch that can pass all tests is found, i.e., \textit{plausible patch}. The whole patch validation is often terminated once finding plausible patches or reaching the time budget. 

Existing work suggests that patch validation is very time consuming~\cite{DBLP:conf/issta/LouGLZZHZ20, DBLP:conf/icse/HuaZWK18/sketchfix}, because (1) the number of generated patches is large,  and (2) each patch validation requires non-trivial time to execute the original tests. \revised{To date, various APR systems have been proposed to generate patches with different strategies}, and they can be categorized into the following categories according to the way of patch generation. 
(1) Heuristic-based APR~\cite{ DBLP:conf/issta/JiangXZGC18}, (e.g., GenProg~\cite{DBLP:journals/tse/YuanB20/arja}, \revised{ARJA-e~\cite{DBLP:journals/tosem/YuanB20}, VarFix~\cite{DBLP:conf/sigsoft/WongSKG21}}) iteratively explore the search space of program modifications via certain heuristics;
(2) Constraint-based APR~\cite{DBLP:conf/icse/XiongWYZH0017/acs, DBLP:conf/ssbse/MartinezM18/cardumen, DBLP:conf/icse/DurieuxM16/dynamoth}, e.g., Nopol~\cite{DBLP:journals/tse/XuanMDCMDBM17/nopol}, generates patches for conditional expressions by transforming them into constraint solving tasks;
(3) Template-based APR~\cite{DBLP:conf/wcre/LiuK0B19/avatar, DBLP:journals/ese/KoyuncuLBKKMT20/fixminer},  e.g., KPar~\cite{DBLP:conf/icst/LiuKB0KT19/kpar},  designs predefined patterns to guide patch generation;
(4) Learning-based APR~\cite{DBLP:conf/issta/LutellierPPLW020/coconut, DBLP:conf/icse/Li0N20/dlfix,chen2019sequencer}, e.g., CocoNut~\cite{DBLP:conf/issta/LutellierPPLW020/coconut},  adopts machine/deep learning techniques to generate patches based on existing code corpus. 

\begin{table}[htb]
    \centering
   \caption{Revisiting RTS strategies on APR systems}\label{table:revisit}
   	\begin{adjustbox}{width=0.8\columnwidth}
    \begin{tabular}{|rrr| rrr| rrr|} 
    \hline
    
    \textbf{APR} & \textbf{Time} & \textbf{RTS}  &   \textbf{APR} & \textbf{Time} & \textbf{RTS}   &   \textbf{APR} & \textbf{Time} & \textbf{RTS} \\ \hline \hline

    PAR~\cite{DBLP:conf/icse/KimNSK13/par} & 2013 & No &
   jGenProg~\cite{DBLP:conf/issta/MartinezM16/astor} &  2016  & No &
   jKali~\cite{DBLP:conf/issta/MartinezM16/astor} &  2016  & No \\

    jMutRepair~\cite{DBLP:conf/issta/MartinezM16/astor} &  2016  & No & DynaMoth~\cite{DBLP:conf/icse/DurieuxM16/dynamoth} &2016 & No 
    &  xPAR~\cite{DBLP:conf/wcre/LeLG16/xpar} & 2016  & No  \\ 
    
    HDRepair~\cite{DBLP:conf/wcre/LeLG16/xpar} & 2016 & No & 
    NPEFix~\cite{DBLP:conf/wcre/DurieuxCSM17/npefix} & 2017  & No &
    ACS~\cite{DBLP:conf/icse/XiongWYZH0017/acs} &2017  & No \\
    
    Genesis~\cite{DBLP:conf/sigsoft/LongAR17/gensis} & 2017 &  No &
    jFix/S3~\cite{DBLP:conf/sigsoft/LeCLGV17/s3} & 2017 &  No &
    EXLIR~\cite{DBLP:conf/kbse/SahaLYP17/elixir} & 2017 &  No  \\
    
    JAID~\cite{DBLP:conf/kbse/Chen0F17/jaid} & 2017 & No &
    ssFix~\cite{DBLP:conf/kbse/XinR17/ssfix} &  2017 & No &   
    SimFix~\cite{DBLP:conf/issta/JiangXZGC18}  & 2018 &  No \\

    Cardumen~\cite{DBLP:conf/ssbse/MartinezM18/cardumen} &  2018  & No &
    \cellcolor{gggray}SketchFix~\cite{DBLP:conf/icse/HuaZWK18/sketchfix} & \cellcolor{gggray}2018  &  \cellcolor{gggray}Stmt  & 
    LSRepair~\cite{DBLP:conf/apsec/LiuKK0B18/LSRepair} & 2018 & No \\

    SOFix~\cite{DBLP:conf/wcre/LiuZ18/sofix} &  2018 &  No &
    \cellcolor{ggggray}\capgen{}~\cite{DBLP:conf/icse/WenCWHC18}  & \cellcolor{ggggray}2018 & \cellcolor{ggggray}Class   & 
      \cellcolor{gggray}ARJA~\cite{DBLP:journals/tse/YuanB20/arja} & \cellcolor{gggray}2018 & \cellcolor{gggray}Stmt  \\ 
    
     \cellcolor{gggray}GenProg-A~\cite{DBLP:journals/tse/YuanB20/arja} & \cellcolor{gggray}2018 & \cellcolor{gggray}Stmt  &
    \cellcolor{gggray} Kali-A~\cite{DBLP:journals/tse/YuanB20/arja} & \cellcolor{gggray} 2018 & \cellcolor{gggray} Stmt   
     &\cellcolor{gggray}RSRepair-A~\cite{DBLP:journals/tse/YuanB20/arja} & \cellcolor{gggray}2018&\cellcolor{gggray} Stmt   \\ 
    
    SequenceR~\cite{chen2019sequencer} & 2019 &  No &
     kPAR~\cite{DBLP:conf/icst/LiuKB0KT19/kpar} & 2019 & No & 
     DeepRepair~\cite{DBLP:conf/wcre/WhiteTMMP19/deeprepair} & 2019  & No \\
    
     \cellcolor{gggray}\prapr{}~\cite{DBLP:conf/issta/GhanbariBZ19}  &\cellcolor{gggray} 2019 & \cellcolor{gggray}Stmt & Hercules~\cite{DBLP:conf/icse/SahaSP19/hercules} & 2019 & No&
     GenPat~\cite{DBLP:conf/kbse/JiangRXZ19/genpat} & 2019& No \\ 
    
     AVATAR~\cite{DBLP:conf/wcre/LiuK0B19/avatar} & 2019& No &
     TBar~\cite{DBLP:conf/issta/LiuK0B19/tbar} & 2019 & No &
     Nopol~\cite{DBLP:journals/tse/XuanMDCMDBM17/nopol} &2019 & No\\  
     ConFix~\cite{DBLP:journals/ese/KimK19/confix} & 2019 & No &
     FixMiner~\cite{DBLP:journals/ese/KoyuncuLBKKMT20/fixminer} & 2020  & No &
     CocoNut~\cite{DBLP:conf/issta/LutellierPPLW020/coconut}  & 2020  & No \\ 
    
      DLFix~\cite{DBLP:conf/icse/Li0N20/dlfix} & 2020 & No 
     &CURE~\cite{DBLP:conf/icse/JiangL021} &2021& No
     &Recoder~\cite{DBLP:conf/sigsoft/ZhuSXZY0Z21} & 2021 & No \\ 
     
     Reward~\cite{DBLP:conf/icse/YeMM22} & 2022 & No & 
     DEAR~\cite{DBLP:conf/icse/Li0N22} & 2022 &No & 
     AlphaRepair~\cite{DBLP:conf/sigsoft/XiaZ22} & 2022 & No \\

   \revised{ARJA-e~\cite{DBLP:journals/tosem/YuanB20}} &  \revised{2020} &  \revised{No}  &
     \revised{VarFix~\cite{DBLP:conf/sigsoft/WongSKG21}} &  \revised{2021}&  \revised{No} &
    && \\ \hline

    \end{tabular}
    \end{adjustbox}
\end{table}


\subsection{RTS in Patch Validation}~\label{sec:revisit}
In addition to the large number of generated patches, each patch also needs to be executed against the original tests, which can be very  costly\Comment{ validation is time consuming due to the non-trivial costs in test executions}.
In fact, the \textit{generate-and-validate} procedure in APR is very similar to the regression testing scenario that patches are modifications of the original buggy program and each of them needs to be validated by the existing test suite. 
Therefore, it is intuitive to adopt regression testing techniques in patch validation, such as improving the repair efficiency by executing only the tests affected by the patch via RTS.
We then systematically revisit RTS techniques adopted in existing APR systems. In particular, we consider APR systems targeting Java  programs due to its popularity in the APR community, and use two sources of information to collect representative APR systems (i.e., \href{http://program-repair.org}{\texttt{program-repair.org}} and the living  review of APR~\cite{livingapr}). For each APR system, we check which RTS strategy is used both in its paper and its implementation (if the source code is released). Table~\ref{table:revisit} summarizes the results of our literature review.\Comment{ In summary, we find that there is no APR system that has adopted any RTP strategy to accelerate patch validation. In particular, RSRepair-A~\cite{DBLP:journals/tse/YuanB20/arja}, the only existing APR system that leverages a fine-level test prioritization strategy in patch validation, executes the tests that fail on more validated patches earlier. Note that such a prioritization strategy has been proposed specifically for the APR scenario~\cite{DBLP:conf/icsm/QiML13}, and does not belong to RTP category. As for RTS strategies, Table~\ref{table:revisit} summarizes the results of our literature review.}

Based on the table, we have the following findings. 
(1) Most existing APR systems, including the latest system proposed in 2022, have not adopted any RTS strategy to optimize the efficiency of patch validation. In other words, the whole test suite is repeatedly executed against each generated patch, including the tests that are not affected by the patch at all. \revised{Although some system (i.e., VarFix) adopts random sampling to reduce the test executions, it still execute the entire test suite after the sampled test are passing. } (2) The few APR systems which adopt RTS are all based on dynamic RTS. This makes sense as prior regression testing studies show that static RTS can be imprecise and unsafe (e.g., due to Java Reflections)~\cite{DBLP:conf/sigsoft/LegunsenHSLZM16,DBLP:journals/pacmpl/ShiHZML19}.
(3) For the few APR systems using RTS, they adopt RTS at different granularities (highlighted by different colors in the table) without clear justification. For example, \prapr{} directly performs dynamic statement-level RTS, while \capgen{} uses class-level RTS.

\Comment{As for RTP strategies, we find that there is no APR system that has adopted any RTP strategy to accelerate patch validation. Several APR systems (e.g., PraPR) execute the originally-failed tests before originally-passed tests with an early-exit mechanism, 
but they do not  further schedule the execution order within the originally-failed/passed tests. In particular, RSRepair-A~\cite{DBLP:journals/tse/YuanB20/arja}, the only existing APR system that leverages a fine-level test prioritization strategy in patch validation, executes the tests that fail on more validated patches earlier. Note that such a prioritization strategy has been proposed specifically for the APR scenario~\cite{DBLP:conf/icsm/QiML13}, and does not belong to RTP category.}


\vspace{5mm}
\begin{mdframed}[linecolor=gray,roundcorner=12pt,backgroundcolor=gray!15,linewidth=3pt,innerleftmargin=2pt, leftmargin=0cm,rightmargin=0cm,topline=false,bottomline=false,rightline = false]
\textbf{Motivation:} RTS is neglected by most APR systems and different RTS strategies are randomly adopted for those few systems using RTS. Therefore, it remains unknown how necessary it is to adopt RTS in APR and how different RTS techniques affect APR efficiency. To bridge this knowledge gap, we perform the first extensive study on the impact of RTS on APR efficiency.
\end{mdframed}

\section{Study Design}

\subsection{Preliminaries}~\label{sec:term} 
We first formally define key conceptions used in this paper.

\subsubsection{Patch Validation Matrix}~\label{sec:matrix}
Given a buggy program $\prog$ and its test suite $\tsts$ (with at least one failed test), an APR system generates a list of candidate patches $\patches{}$ after the  patch generation phase. 
 
\begin{theorem}\label{def:patchmatrix}
\textbf{Patch validation matrix} $\pMatrix$ defines the execution results of all tests on all patches. In particular, each cell $\pMatrixCell{\patch}{\tst}$ represents the execution result of test $\tst\in\tsts$ on patch $\patch\in\patches$, which can have the following values: (1) \skips, if $\tst$ has not been selected by RTS and thus skipped for execution, (2) \fail, if $\tst$ fails on the patch $\patch$, (3) \pass, if $\tst$ passes on the patch $\patch$, and (4) \unknown{}, if $\tst$ is selected by RTS but has not been executed\Comment{ before the validation for $\patch$ terminates} (i.e., its execution result remains unknown).
\end{theorem}

The early-exit mechanism is a common efficiency optimization widely enabled in existing APR systems~\cite{DBLP:conf/issta/JiangXZGC18, DBLP:conf/issta/GhanbariBZ19}, which stops validation for the current patch once it fails on any test. 
In this scenario, there are many tests whose results remain unknown, resulting in a validation matrix with ``\unknown{}'' cells. We denote such a matrix as a \textbf{partial patch validation matrix} $\pMatrixpart$. 
Meanwhile, in the other scenario where the execution results of all tests are required~\cite{benton2020effectiveness, DBLP:conf/issta/LouGLZZHZ20}, the early-exit mechanism is often disabled. 
In this scenario, all the selected tests would be executed even when there are some tests that already failed during patch validation, leaving no ``\unknown{}'' cell in the matrix $\pMatrix$. We denote such a matrix as a \textbf{full patch validation matrix} $\pMatrixfull$. 
The following are examples for full and partial matrices when there is no RTS adopted in patch validation. For example, patch validation for $\patch_1$ stops by the failure of $\tst_1$, and thus the execution results of $\tst_2$ and $\tst_3$ on $\patch_1$ remain unknown in $\pMatrixpart$.

\begin{align} \label{eq:m1}\scriptsize
    \pMatrixfull =  \begin{bmatrix} 
    & \tst_1 & \tst_2 & \tst_3   \\ 
    \patch_1 &  $\fail$ & $\pass$ & $\pass$  \\    
    \patch_2 &  $\fail$ & $\fail$ & $\pass$ \\  
    \patch_3 &  $\pass$ & $\pass$ & $\pass$ \\  
    \end{bmatrix} 
    & & \scriptsize
    \pMatrixpart =  \begin{bmatrix} 
    & \tst_1 & \tst_2 & \tst_3  \\ 
    \patch_1 &  $\fail$ & $\unknown$ & $\unknown$  \\    
    \patch_2 &  $\fail$ & $\unknown$ & $\unknown$ \\  
    \patch_3 &  $\pass$ & $\pass$ & $\pass$\\  
    \end{bmatrix} 
\end{align}


\subsubsection{Studied RTS Strategies}~\label{sec:techs}
Based on the literature review, we focus on dynamic RTS strategies that utilize test coverage in buggy programs as test dependencies, since static RTS can be imprecise and unsafe~\cite{DBLP:conf/sigsoft/LegunsenHSLZM16,DBLP:journals/pacmpl/ShiHZML19}. Given a candidate patch $\patch$, $\modcode$ denotes the set of code elements modified by $\patch$,  and $\covcode$ denotes the set of code elements in $\prog$ that are covered by the test $\tst$.

\begin{theorem}
\label{def:aprrts}
Given a patch $\patch$ and the whole test suite $\tsts$, a \textbf{dynamic RTS strategy for APR} selects a subset of tests $\tstsrts$ for execution. In particular, for precise and safe RTS, each test in $\tstsrts$ should cover at least one modified code element, i.e., $ \forall{}\tst{}\in\tstsrts{}, \modcode{} \cap \covcode \neq{}\emptyset{}$, while each test not in $\tstsrts$ should not cover any modified element, i.e., $ \forall{}\tst{}\notin\tstsrts{},  \modcode{} \cap {}\covcode = \emptyset{}.$

\end{theorem}

\Comment{Such a dynamic RTS strategy for APR is safe since it selects all the tests that are affected by the changed code; meanwhile it is quite precise since it selects only the tests that are affected by the changed code.}Note that the above definition is simplified for the ease of understanding. Our actual implementation for ensuring RTS precision and safety is actually more complicated and handles all types of Java code changes (e.g., method-overriding hierarchy changes) following prior safe RTS work~\cite{DBLP:conf/icse/Zhang18, DBLP:conf/issta/GligoricEM15}.
According to the granularity of modified code elements, we study three RTS strategies: class-level (\rtsclass{}), method-level (\rtsmethod{}), and statement-level RTS (\rtsstate{}). In addition, we regard no RTS adoption as the baseline test  selection strategy, denoted as \norts{}. 
In the example, if $\tst_1$ and $\tst_3$ cover the statements modified by $\patch_1$, while $\tst_1$ and $\tst_2$ cover the statements modified by $\patch_2$ and $\patch_3$, 
the partial and full validation matrices associated with \rtsstate{} can be represented as Equation~\ref{eq:m2}. 
For example, since $\tst_3$ does not cover any statement modified by $\patch_3$, \rtsstate{} skips $\tst_3$ in the validation for $\patch_3$.
However, if $\tst_3$ happens to cover other statements in the same class as the modified statements (i.e., $\tst_3$ covers the classes modified by $\patch_3$), \rtsclass{} would select $\tst_3$ when validating $\patch_3$.
In this way, RTS at coarser granularities tend to select more tests.

\begin{align} \label{eq:m2}
\vspace{1mm}\scriptsize
    \pMatrixfull =  \begin{bmatrix} 
    & \tst_1 & \tst_2 & \tst_3   \\ 
    \patch_1 &  $\fail$ & $\skips$ & $\pass$  \\    
    \patch_2 &  $\fail$ & $\fail$ & $\skips$ \\  
    \patch_3 &  $\pass$ & $\pass$ & $\skips$ \\  
    \end{bmatrix} 
    & & \scriptsize
    \pMatrixpart =  \begin{bmatrix} 
    & \tst_1 & \tst_2 & \tst_3   \\ 
    \patch_1 &  $\fail$ & $\skips$ & $\unknown$  \\    
    \patch_2 &  $\fail$ & $\unknown$ & $\skips$ \\  
    \patch_3 &  $\pass$ & $\pass$ & $\skips$ \\  
    \end{bmatrix}
\end{align}

\subsubsection{Efficiency Measurement}
Recent work~\cite{DBLP:conf/icse/0001WKKB0WKMT20} on APR efficiency adopts the number of patches as the efficiency metric. However, in our work, simply counting the number of patches could be imprecise, because it treats each patch as equally costly and measures repair efficiency in an oversimplified way. 
For example, given $\pMatrixpart{}$ in Equation~\ref{eq:m1}, both $\patch_2$ and $\patch_3$ would be regarded as one program execution, even if $\patch_3$ actually executes more tests than $\patch_2$. Therefore, in this study, we define the following metrics based on the number or time of test executions for precise efficiency measurement.

\begin{theorem}~\label{def:count}
Given a patch validation matrix $\pMatrix{}$, its \textbf{accumulated number of test executions} $NT_{num}(\pMatrix{})$, sums up the number of test executions on all the validated patches, i.e., $NT_{num}(\pMatrix{}) = \sum{}{}{\pMatrixCell{\patch}{\tst}}$, if $\  \pMatrixCell{\patch}{\tst} \neq{}$\skips{}$ \ \land \  \pMatrixCell{\patch}{\tst}\neq{}$ \unknown; its \textbf{accumulated time of test executions} $NT_{time}(\pMatrix{})$, sums up the time of test executions on all the validated patches, i.e., $NT_{time}(\pMatrix{}) = \sum{}{}{f_t(\pMatrixCell{\patch}{\tst}})$, if $\  \pMatrixCell{\patch}{\tst} \neq{}$\skips{}$ \ \land \  \pMatrixCell{\patch}{\tst}\neq{}$ \unknown, and the function $f_t$ returns the execution time of the given test $\tst$ on the patch $\patch$.
\end{theorem}

\begin{theorem}~\label{def:ratio}
Given the patch validation matrix $\pMatrix{}$ without regression testing technique, and the matrix $\rtsMatrix{}$ generated by certain regression testing strategy, $Reduction_{num/time}$ measures the efficiency improvement achieved by the RTS strategy compared to no RTS, i.e., $Reduction_{num/time}$ $= \frac{NT_{num/time}(\pMatrix{}) - NT_{num/time}(\rtsMatrix{})}{NT_{num/time}(\pMatrix{})}$.
\end{theorem}

For example, without any RTS (in Equation~\ref{eq:m1}), $NT_{num}(\pMatrixfull{})$ is 9 and $NT_{num}(\pMatrixpart{})$ is 5. Meanwhile, with \rtsstate{} (in Equation~\ref{eq:m2}), $NT_{num}(\pMatrix{}_f^{RTS})$ is 6 and $NT_{num}(\pMatrix{}_p^{RTS})$ is 4. \rtsstate{} achieves 33.33\% /20.00\% $Reduction_{num}$ in the full/partial matrix. Such efficiency difference cannot be captured by the number of patches. For example, if only considering the number of patches in $\pMatrix{}_p^{RTS}$ and $\pMatrix{}_p$, there is no reduction achieved by RTS at all.


In this work, we mainly presents the results on the number of test executions. In fact, we find that results between the number or the time of test executions are consistent (more details in Section~\ref{sec:dis}). While execution time is often dependent on many factors (e.g., specific implementations and test execution engines) unrelated to APR approaches~\cite{DBLP:conf/icse/0001WKKB0WKMT20}, the number of test executions is more stable and more suitable for future work to reproduce. In addition, we also discuss the RTS impact on repair effectiveness in Section~\ref{sec:dis}.

\subsection{Research Questions} \label{sec:rq}

\begin{itemize}

    \item \textbf{RQ1: Revisiting APR efficiency.} We revisit the efficiency of APR systems by making the first attempt to (1) measure repair efficiency by the number of test executions, and (2) compare the efficiency of different APR systems with consistent RTS configurations to eliminate the bias from RTS strategies.

    \item \textbf{RQ2: Overall impact of RTS strategies.} We investigate the efficiency improvement achieved by different RTS strategies based on their reduction of test executions.
    
    \item \textbf{RQ3: Impact on different patches.} We further study the impact of RTS on patches of different characteristics (i.e., fixing capabilities and
fixing scopes) to find out what kinds of patches are more susceptible to  RTS.


\Comment{ In particular, we categorize patches as follows:
  \begin{itemize}[leftmargin=10pt, itemindent=0pt, topsep=0pt]
        \item \textbf{RQ3a: Impact on patches of different fixing capabilities.} We categorize patches according to their fixing capabilities, i.e., the patches that can pass more tests are considered to have stronger fixing capabilities. We then study the impact of RTS strategies on these patches with different fixing capabilities.  
        
        \item \textbf{RQ3b: Impact on patches of different fixing scopes.} We categorize patches according to their scopes of patched code, i.e., the number of code elements modified by the patches. We then study the impact of RTS strategies on such patches.  
    \end{itemize}}
    
    \item \textbf{RQ4: Impact with the full patch validation matrix.} In RQ1-RQ3, we investigate APR efficiency with the default \emph{partial} patch validation matrix (with early-exit), which is widely adopted in many APR systems for the sake of efficiency. In this RQ, we further study the impact of RTS strategies with the \emph{full} validation matrix (without early-exit).\Comment{ It is the common setting when execution results of all tests are required, e.g., in unified debugging~\cite{DBLP:conf/issta/LouGLZZHZ20, benton2020effectiveness}. }
    \item \textbf{RQ5: Combining test selection with prioritization.} In addition to RTS strategies, when the early-exit is enabled, the execution order of tests jointly decides the number of test executions. Therefore, we further combine the studied RTS strategies with state-of-the-art test prioritization techniques, and investigate their impact on APR efficiency.
\Comment{    \item \textbf{RQ5: Joint impact of test selection and prioritization.} 
    In this RQ,  we further combine the studied RTS strategies with state-of-the-art test prioritization techniques\lingming{~\cite{trp, tot, additional}} to investigate their joint impact on APR efficiency.}
\end{itemize}

\subsection{Benchmark}
We perform our study on the benchmark Defects4J V.1.2.0 ~\cite{DBLP:conf/issta/JustJE14}, which is the version used most widely by existing APR work~\cite{DBLP:conf/icse/XiongWYZH0017/acs,DBLP:conf/issta/LiuK0B19/tbar,DBLP:conf/issta/GhanbariBZ19}. Defects4J V.1.2.0 includes 395 real-world bugs from six real-world software systems. 
\revised{We choose Defecst4J, as (1) some studied APR systems are only implemented for Java (e.g., SimFix), and (2) most studied systems are evaluated on Defects4J in their previous work while using a consistent widely-used benchmark can better position our findings in the domain.}


\subsection{Studied APR Systems}
Following the recent studies on APR ~\cite{DBLP:conf/icse/0001WKKB0WKMT20, benton2020effectiveness}, our study selects APR systems for Java program according to the following criteria. (1) The source code of the APR system is publicly available because we need to modify patch validation settings.  (2) The APR systems can be successfully applied to Defects4J subjects. (3) The APR systems require no extra input data besides program source code and the available test suite. 
In this way, we successfully collect 16 APR systems fulfilling the requirements above, and all of them are also used as the studied APR systems in previous work~\cite{DBLP:conf/icse/0001WKKB0WKMT20, benton2020effectiveness}. 
However, these APR systems use test cases at different granularities: the Astor family~\cite{DBLP:conf/issta/MartinezM16/astor} (including \jmutrepair{}, \jgenprog{}, \jKali{}) and \cardumen{}~\cite{DBLP:conf/ssbse/MartinezM18/cardumen} regard each test class as a test case, while the remaining systems all regard each test method as a test case.
Since RTS results at different test-case granularities are incomparable, we use the more popular test-method granularity with 12 APR systems in our study, including (1) \textit{constraint-based systems}:  \acs{} and \dynamoth{}, (2) \textit{heuristic-based systems}: \arja, \genprog{},  \kali{}, RSRepair-A, and \simfix{}, (3)  \textit{template-based systems}: \avatar{}, \fixminer{},  \kpar{}, \prapr{}, and \tbar{}. In total, \emph{the studied APR systems generate 2,532,915 patches for all the Defects4J subjects studied, while each patch can involve up to thousands of test executions.}




\subsection{Implementation Details}
To implement studied RTS strategies, following state-of-the-art Ekstazi~\cite{DBLP:conf/issta/GligoricEM15} and HyRTS~\cite{DBLP:conf/icse/Zhang18}, we leverage on-the-fly bytecode instrumentation with ASM~\cite{ASM} and Java Agent~\cite{javaagent} for lightweight test dependency collection at the class/method/statement level on buggy programs.\Comment{
, and leverage smart CheckSums~\cite{DBLP:conf/issta/GligoricEM15} for fast change computation.} 
For all studied APR systems, we modify their source code to collect patch/test execution results with different RTS and validation settings. 
\revised{We only keep the compilable patches (85\% compilation rate) in our experiments as RTS strategies cannot be applied to un-executable patches.}
For each system, we try up to three re-executions for failed/error repair attempts. In all RQs, we schedule all originally-failed tests prior to originally-passed tests, which is also the common practice of all recent APR systems because the former are more likely to fail again on patches.
\revised{For each APR system, we mainly follow the same configurations (such as timeout and JDK version) as its original publication.}
We manually modified all studied APR systems to collect the detailed patch execution information required by unified debugging and ensured our modified versions did not impact underlying tool functionality. Each system used original time settings suggested by the original papers. Each tool was executed using the same JDK version found in the tool’s original publication, allowing us to obtain repair execution results as close as possible to the tool’s original results.

From RQ1 to RQ4, within the originally-failed tests or the originally-passed tests, we prioritize them by their default lexicographical order to eliminate the impact of different test execution orders. Our data is available at~\cite{ourwebsite}, including patch results and impact on repair effectiveness of all studied APR tools.

\subsection{Threat to Validity}
Our study focuses on APR systems for Java programs, which might threaten generality of our findings. We mitigate this threat by performing our study on a large spectrum of APR systems belonging to different categories and a widely-used benchmark with hundreds of real bugs~\cite{DBLP:conf/issta/GhanbariBZ19, DBLP:conf/icse/WenCWHC18, DBLP:conf/issta/JiangXZGC18}. In addition, to mitigate the threat in faulty implementations, we carefully review our RTS implementations and experimental scripts; the source code of all the studied APR systems was directly obtained from their authors or original open-source repositories; we further ensured that each APR system performs consistently before and after our modifications.

\section{Results analysis}
\begin{table*}[htb]
    \centering
   \caption{Number of test executions among  APR systems}\label{table:rq0_cmpall}

   	\begin{adjustbox}{width=0.97\columnwidth}
    \begin{tabular}{|r|l||rrrrrrrrrrrr|}

    \hline
    
    \textbf{Subject} &  \textbf{RTS Strategy} &  \textbf{\prapr{}} &  \textbf{\simfix{}} &  \textbf{\avatar{}} &  \textbf{\kpar{}} &  \textbf{\tbar{}} &  \textbf{\fixminer{}} &  \textbf{\dynamoth{}} &  \textbf{\acs{}} &  \textbf{\arja{}} &  \textbf{\kali{}} &  \textbf{\genprog{}} &  \textbf{\rsrepair{}} \\ \hline \hline
    
    \multirow{5}{*}{Lang} & \#Patch&381.27& 275.98 &4.63&4.79&4.76&4.56&0.16&1.00&522.97&26.92&628.85&289.23\\
 & \#Test (\norts{})&2,117.61& 8,802.36 &321.41&163.19&437.52&233.15&1,256.86&1,140.80&68,448.64&799.43&86,910.25&10,945.44\\
 & \#Test (\rtsclass{})&447.31& 291.91 &19.48&18.30&24.63&18.88&25.71&64.40&1,571.25&59.54&2,746.19&620.69\\
 & \#Test (\rtsmethod{})&414.15& 244.86 &12.11&11.89&12.63&11.88&2.86&1.80&1,196.83&51.54&1,376.81&556.39\\
 & \#Test (\rtsstate{})&412.59& 244.79 &11.44&11.59&11.96&11.50&2.00&1.20&1,176.53&50.63&1,369.64&534.58\\ \hline
\multirow{5}{*}{Math} & \#Patch&1,483.12& 469.45 &5.84&5.12&5.01&4.76&0.23&1.00&804.09&10.11&858.19&411.58\\
 & \#Test (\norts{})&11,879.77&  13,723.01 &168.78&237.05&640.65&239.38&2,042.74&3,071.33&104,804.90&705.64&73,174.10&9,661.55\\
 & \#Test (\rtsclass{})&1,711.16& 679.05 &12.92&46.66&34.66&9.47&50.79&241.33&4,828.21&22.00&5,231.08&828.93\\
 & \#Test (\rtsmethod{})&1,573.21& 403.17 &8.20&7.25&8.18&7.03&32.53&14.83&2,105.64&15.97&1,610.38&498.28\\
 & \#Test (\rtsstate{})&1,557.41& 379.67 &7.81&6.86&7.25&6.23&27.00&6.33&1,382.11&14.10&1,419.85&463.63\\ \hline
\multirow{5}{*}{Time} & \#Patch&1,466.54&396.50&2.50&2.00&2.23&1.08&0.23&1.00&335.92&13.23&380.62&107.46\\
 & \#Test (\norts{})&7,782.19& 3,754.42
 &1,065.00&4.73&359.09&490.00&1,303.67&3,894.00&34,310.82&724.00&141,456.00&2,652.50\\
 & \#Test (\rtsclass{})&2,513.88& 658.67
 &232.55&4.73&80.82&107.38&537.33&2,042.00&7,337.91&325.70&29,401.40&635.00\\
 & \#Test (\rtsmethod{})&1,848.62& 372.08
 &9.73&4.73&6.55&5.25&8.67&52.00&3,822.82&55.30&10,596.50&413.90\\
 & \#Test (\rtsstate{})&1,808.12& 368.08
 &9.73&4.73&6.55&5.25&8.67&27.00&1,781.91&38.50&1,903.60&291.50\\ \hline
\multirow{5}{*}{Chart} & \#Patch&784.88&588.21
&5.00&4.56&4.40&4.48&0.40& - &627.76&42.36&729.72&333.16\\
 & \#Test (\norts{})&5,517.96& 2,459.42
 &272.84&107.61&367.74&102.12&1,640.80&-&202,236.87&2,035.27&104,090.08&9,552.46\\
 & \#Test (\rtsclass{})&1,053.80& 529.25
 &8.26&8.00&8.79&7.12&77.30&-&27,627.48&133.45&5,694.38&758.38\\
 & \#Test (\rtsmethod{})&868.24& 485.96
 &6.89&6.89&6.37&6.59&16.70&-&6,853.48&61.45&2,645.12&450.67\\
 & \#Test (\rtsstate{})&863.96& 485.92
 &6.89&6.89&6.37&6.59&16.30&-&2,356.57&56.82&1,746.42&386.71\\ \hline
\multirow{5}{*}{Closure} & \#Patch&14,725.00& 511.03
&2.96&2.61&2.94&0.42&-&-&-&-&-&-\\
 & \#Test (\norts{})&86,651.19& 7,995.31
 &342.78&378.73&478.09&2.89&-&-&-&-&-&-\\
 & \#Test (\rtsclass{})&47,374.48& 3,229.28
 &140.81&154.94&197.02&2.89&-&-&-&-&-&-\\
 & \#Test (\rtsmethod{})&28,476.71&922.37
 &6.17&9.90&6.67&2.89&-&-&-&-&-&-\\
 & \#Test (\rtsstate{})&21,825.07& 662.63
 &4.49&3.92&4.40&2.89&-&-&-&-&-&-\\ \hline
\multirow{5}{*}{Mockito} & \#Patch&2,307.92& - &1.67&2.00&1.81&1.67&-&-&-&-&-&-\\
 & \#Test (\norts{})&3,753.19&-&267.75&41.27&200.27&4.62&-&-&-&-&-&-\\
 & \#Test (\rtsclass{})&2,857.81&-&97.56&26.00&146.47&4.62&-&-&-&-&-&-\\
 & \#Test (\rtsmethod{})&2,555.61&-&7.62&5.80&48.00&4.62&-&-&-&-&-&-\\
 & \#Test (\rtsstate{})&2,547.64&-&7.38&5.80&7.33&4.62&-&-&-&-&-&-\\ \hline
    \end{tabular}
    \end{adjustbox}
    
\end{table*} 

\subsection{RQ1: Revisiting APR Efficiency} ~\label{sec:rq0}
Table~\ref{table:rq0_cmpall} presents the number of test executions with different RTS strategies. 
We also present the number of validated patches in the table \revised{(i.e., \#Patch is the average patch number of each bug)}. Note that cells are empty when there is no valid patch generated. Based on the table, we have the following observations.

First, measuring APR efficiency by a more precise metric (i.e., the number of test executions) can draw totally different conclusions from the number of validated patches advocated by prior work~\cite{DBLP:conf/icse/0001WKKB0WKMT20}.
For example, for each buggy version of Math, \simfix{} (469.45 patches) validates many more patches than \rsrepair{} (411.58 patches), showing that \simfix{} is less efficient than \rsrepair{} in terms of the number of validated patches. 
Such a conclusion is consistent with the finding in  previous work~\cite{DBLP:conf/icse/0001WKKB0WKMT20}, which also measures APR efficiency by the number of validated patches. 
However, the conclusion can be opposite if these systems are compared by the number of test executions. For example, with class-level RTS, \simfix{} executes 679.05 tests and \rsrepair{} executes 828.93 tests per buggy version of Math, indicating that \simfix{} is more efficient than \rsrepair{}. The reason is that although \rsrepair{} generates fewer patches than \simfix{}, each patch of \rsrepair{} executes many more tests than \simfix{}. Such inconsistencies between the number of test executions and patches are prevalent. Hence, it is imprecise to measure APR efficiency by simply counting the number of patches, since different patches can execute totally different number of tests. Future work on APR efficiency should also consider the number of test executions.

Second, comparing APR efficiency with different RTS strategies can also deliver opposite conclusions.  
For example, on subject Time, \rsrepair{} executes 413.90 and 291.50 tests with \rtsmethod{} and  \rtsstate{}, while \simfix{} executes 372.08 and 369.08 tests with \rtsmethod{} and \rtsstate{}.
When both systems are configured with method-level RTS, \rsrepair{} is considered as less efficient than \simfix{}; however, when they are compared  under statement-level RTS,  \rsrepair{} is actually more efficient than \simfix{}. 
Moreover, if APR systems are compared under different RTS strategies respectively, the conclusion may further be skewed.
Therefore, it is important to adopt the same RTS strategy when comparing the repair efficiency among different APR systems. 
We also strongly recommend the future APR work to explicitly describe their  RTS strategy adopted in experiments, so that their follow-up work can mitigate the threats in RTS configuration. 

\finding{\textit{\revised{using inconsistent RTS configurations can skew the conclusion of APR efficiency.} Future APR work should explicitly describe their adopted RTS strategies and the efficiency comparison among multiple APR systems should guarantee a consistent RTS configuration. }}

\Comment{\finding{\textit{Compared to a more precise metric, the number of patches used before can have misleading conclusions on APR efficiency.
RTS strategy is an important configuration associated with APR efficiency measurement. Future work should explicitly describe the RTS strategy used in experiments and the efficiency comparison among APR systems should guarantee a consistent RTS configuration in each APR system.}}}

\subsection{RQ2: Overall Impact of RTS Strategies} ~\label{sec:rq1}
Table~\ref{table:rq1_cmpnorts_ratio} presents the reduction of the number of test executions achieved by each RTS strategy compared to no RTS. 
Based on the table, for all the studied APR systems on almost all the subjects, adopting RTS can remarkably reduce the number of test executions in patch validation.
For example, on the subject Math, \dynamoth{} skips 93.33\% of test executions if it adopts \rtsstate{}.
Even for RTS at the coarsest level (i.e., \rtsclass{}), the average reduction ranges from 4.66\% to 66.93\% among different APR systems. Note that at some cases there is no difference among RTS strategies, e.g., Fixminer on Closure, because all generated patches fail at the first executed test (i.e., the first originally-failed test still fails on these patches). We would further analyze the individual impact on different patches in Section~\ref{sec:rq2}.
In summary, our results suggest that RTS can significantly improve the efficiency of patch validation and future APR work should no longer overlook such an important optimization.

\finding{\textit{for all studied APR systems, adopting RTS can remarkably improve APR efficiency and should be considered in future APR work.}}



Overall, there is a common trend that \rtsstate{} always executes the least tests while \rtsclass{} exhibits the most tests. This is not surprising because class-level coverage is the coarsest selection criterion, based on which more tests would inherently be selected. In addition, \rtsmethod{}/\rtsstate{} can significantly reduce more test executions (4.36\%/4.69\% on average) than  \rtsclass{}. Such a difference is statistically significant according to the Wilcoxon Signed-Rank Test~\cite{wilcoxon1992individual} at the significance level of 0.05 on almost all the APR systems (i.e., 11 out of 12). Meanwhile, interestingly, the difference between \rtsmethod{} and \rtsstate{} is often subtle. For example, the average difference between \rtsmethod{} and \rtsstate{} is only 0.33\%. In fact, such observation is consistent with prior work in traditional regression testing~\cite{DBLP:conf/icse/Zhang18}, which shows that method-level RTS outperforms class-level RTS but is often not significantly worse than statement-level RTS. The reason is that modern system design principles recommend writing simple method bodies for the ease of maintenance, making the
majority of method body changes directly affect all tests covering the methods (i.e., \rtsstate{} is close to \rtsmethod{}).
Hence, method- and statement-level RTS are more recommended for APR.

\finding{\textit{different RTS strategies exhibit different reduction performance. \rtsstate{} and \rtsmethod{} perform similarly, but both of them significantly outperform \rtsclass{}\Comment{, which in turn substantially outperforms \norts{}}.}}

\begin{table*}
    \centering
   \caption{Reduction of the number of test executions by different RTS strategies}\label{table:rq1_cmpnorts_ratio}

   	\begin{adjustbox}{width=0.98\columnwidth}
    \begin{tabular}{|r|r||rrrrrrrrrrrr|}
    
    \hline
    
    \textbf{Subject} &  \textbf{RTS Strategy} &  \textbf{\prapr{}} &  \textbf{\simfix{}} &  \textbf{\avatar{}} &  \textbf{\kpar{}} &  \textbf{\tbar{}} &  \textbf{\fixminer{}} &  \textbf{\dynamoth{}} &  \textbf{\acs{}} &  \textbf{\arja{}} &  \textbf{\kali{}} &  \textbf{\genprog{}} &  \textbf{\rsrepair{}} \\ \hline \hline
    
\multirow{3}{*}{Lang} & \rtsclass{}&33.33\%&35.66\%&17.92\%&10.57\%&21.48\%&14.80\%&69.99\%&56.37\%&36.67\%&22.16\%&26.11\%&30.12\%\\
& \rtsmethod{}&34.51\%&36.31\%&18.35\%&10.95\%&22.05\%&15.22\%&71.31\%&59.94\%&38.16\%&22.50\%&26.68\%&31.17\%\\
& \rtsstate{}&34.55\%&36.31\%&18.40\%&10.99\%&22.09\%&15.26\%&71.36\%&59.97\%&38.17\%&22.52\%&26.70\%&31.20\%\\
\hline\multirow{3}{*}{Math} & \rtsclass{}&48.91\%&52.45\%&5.46\%&5.36\%&19.91\%&8.63\%&92.17\%&90.59\%&38.66\%&16.44\%&28.77\%&31.73\%\\
& \rtsmethod{}&50.41\%&55.81\%&6.27\%&6.30\%&21.08\%&8.99\%&93.06\%&98.86\%&40.16\%&16.79\%&30.16\%&32.44\%\\
& \rtsstate{}&50.50\%&55.89\%&6.28\%&6.31\%&21.11\%&9.05\%&93.33\%&99.63\%&40.42\%&16.85\%&30.35\%&32.68\%\\
\hline\multirow{3}{*}{Time} & \rtsclass{}&31.25\%&29.42\%&7.14\%&0.00\%&7.15\%&9.83\%&19.62\%&47.56\%&19.59\%&17.90\%&7.97\%&16.26\%\\
& \rtsmethod{}&40.12\%&35.60\%&9.06\%&0.00\%&9.06\%&12.45\%&33.15\%&98.66\%&21.41\%&28.27\%&9.31\%&17.38\%\\
& \rtsstate{}&40.31\%&35.60\%&9.06\%&0.00\%&9.06\%&12.45\%&33.15\%&99.31\%&22.60\%&29.01\%&9.93\%&18.01\%\\
\hline\multirow{3}{*}{Chart} & \rtsclass{}&58.50\%&33.16\%&15.68\%&5.49\%&20.88\%&5.85\%&85.96\%&-&53.39\%&40.87\%&34.93\%&39.33\%\\
& \rtsmethod{}&62.45\%&34.67\%&15.76\%&5.55\%&21.02\%&5.88\%&89.10\%&-&59.10\%&43.78\%&38.81\%&42.02\%\\
& \rtsstate{}&62.52\%&34.67\%&15.76\%&5.55\%&21.02\%&5.88\%&89.12\%&-&59.67\%&44.10\%&39.22\%&42.45\%\\
\hline\multirow{3}{*}{Closure} & \rtsclass{}&31.73\%&20.27\%&3.22\%&3.79\%&3.90\%&0.00\%&-&-&-&-&-&-\\
& \rtsmethod{}&46.18\%&30.51\%&5.28\%&6.56\%&7.38\%&0.00\%&-&-&-&-&-&-\\
& \rtsstate{}&51.80\%&32.01\%&5.30\%&6.65\%&7.42\%&0.00\%&-&-&-&-&-&-\\
\hline\multirow{3}{*}{Mockito} & \rtsclass{}&18.03\%&-&21.10\%&2.77\%&5.49\%&0.00\%&-&-&-&-&-&-\\
& \rtsmethod{}&21.17\%&-&30.69\%&6.43\%&14.82\%&0.00\%&-&-&-&-&-&-\\
& \rtsstate{}&21.30\%&-&30.72\%&6.43\%&19.67\%&0.00\%&-&-&-&-&-&-\\
\hline\multirow{3}{*}{Average} & \rtsclass{}&36.96\%&34.19\%&11.75\%&4.66\%&13.13\%&6.52\%&66.93\%&64.84\%&37.08\%&24.34\%&24.45\%&29.36\%\\
&\rtsmethod{}&42.47\%&38.58\%&14.23\%&5.96\%&15.90\%&7.09\%&71.66\%&85.82\%&39.71\%&27.83\%&26.24\%&30.75\%\\
&\rtsstate{}&43.50\%&38.90\%&14.25\%&5.99\%&16.73\%&7.11\%&71.74\%&86.30\%&40.22\%&28.12\%&26.55\%&31.09\%\\
\hline
    \end{tabular}
    \end{adjustbox}
    
\end{table*}

\subsection{RQ3: Impact on Different Patches}~\label{sec:rq2}
We now investigate the impact of RTS strategies on the patches of different characteristics. Based on the intuition that a patch involving more test executions might be more sensitive to RTS strategies, we categorize patches by their fixing capabilities or fixing code scopes. Section~\ref{sec:cate1} and Section~\ref{sec:cate2} present the impact on these patch categorizations respectively.

\begin{table}[htb]
    \centering
   \caption{Reduction on \stopppatch{} and \plausiblepatch{}\Comment{\lingming{why ACS does not have number for Patchp? need to explain in text} \yiling{DONE}}}\label{table:rq2_pp}

   	\begin{adjustbox}{width=0.6\columnwidth}

    \begin{tabular}{|r|| rrr| rrr|}\hline
    
    \multirow{2}{*}{APR} & \multicolumn{3}{c|}{\stopppatch{}} & \multicolumn{3}{c|}{\plausiblepatch{}}  \\ \cline{2-7}
    & \rtsclass{} & \rtsmethod{} & \rtsstate{} & \rtsclass{} & \rtsmethod{} & \rtsstate{}\\ \hline \hline
    
PraPR&73.24\%&85.74\%&87.95\%&87.33\%&97.18\%&97.88\%\\
SimFix&82.88\%&99.11\%&99.15\%&84.50\%&96.77\%&98.22\%\\
AVATAR&81.16\%&98.95\%&99.17\%&82.27\%&99.36\%&99.53\%\\
kPar&66.33\%&98.07\%&98.58\%&83.47\%&99.31\%&99.80\%\\
TBar&68.34\%&98.63\%&99.13\%&76.07\%&95.39\%&99.55\%\\
FixMiner&85.02\%&96.90\%&99.14\%&93.28\%&99.71\%&99.80\%\\
Dynamoth&94.75\%&96.71\%&97.22\%&87.50\%&99.19\%&99.27\%\\
ACS&-&-&-&98.34\%&99.66\%&99.95\%\\
Arja&90.09\%&95.56\%&97.28\%&93.58\%&99.15\%&99.44\%\\
Kali&88.76\%&97.97\%&99.30\%&84.30\%&99.00\%&99.59\%\\
GenProg&88.85\%&96.84\%&99.05\%&94.28\%&99.33\%&99.57\%\\
RSRepair&92.83\%&97.94\%&99.42\%&95.79\%&99.11\%&99.58\%\\

 \hline
 
    \end{tabular}
    \end{adjustbox}

\end{table}

\subsubsection{Impact on patches of different fixing capabilities} \label{sec:cate1}

Based on test execution results, we can categorize patches into groups of different fixing capabilities. (1) \stopfpatch{}: a patch cannot fix all originally-failed tests and thus its validation gets aborted by some originally-failed test; (2) \stopppatch{}: a patch can fix all originally-failed tests but fails on some originally-passed test, and thus its validation gets aborted by some originally-passed test; (3) \plausiblepatch{}: a patch that can pass all originally-failed and originally-passed tests (i.e., plausible patch). Obviously, \stopfpatch{} has the weakest fixing capability and its validation halts extremely early, and thus adopting RTS would not make any difference on its number of test executions. Therefore, we present the impact of RTS strategies on the other two patch groups, i.e., \stopppatch{} and \plausiblepatch{} in Table~\ref{table:rq2_pp}. In particular, each cell shows the reduction ratio of test executions compared to no RTS. Note that since \acs{} generates only \stopfpatch{} and \plausiblepatch{},  its corresponding cells in \stopppatch{} are empty.
As suggested by the table, compared to the overall reduction on all patches (i.e., the reduction in Table~\ref{table:rq1_cmpnorts_ratio}), the impact of RTS strategies on \plausiblepatch{} and \stopppatch{} is even more remarkable. For example, the reduction ratio ranges from 66.33\% to 99.42\% on \stopppatch{}, and ranges from 76.07\% to 99.95\% on \plausiblepatch{}. In addition, compared to all patches, the difference between RTS strategies is also enlarged on \plausiblepatch{} and \stopppatch{}. In summary, RTS can achieve larger reductions for patches with stronger fixing capabilities.

\finding{\textit{the impact of RTS strategies is even more prominent on \plausiblepatch{} and \stopppatch{}, where an  extremely large portion of tests (e.g., up to 99.95\%) get reduced.}}

\begin{figure}[htb]
    \centering
    \includegraphics[width=7.5 cm, height= 3.8 cm]{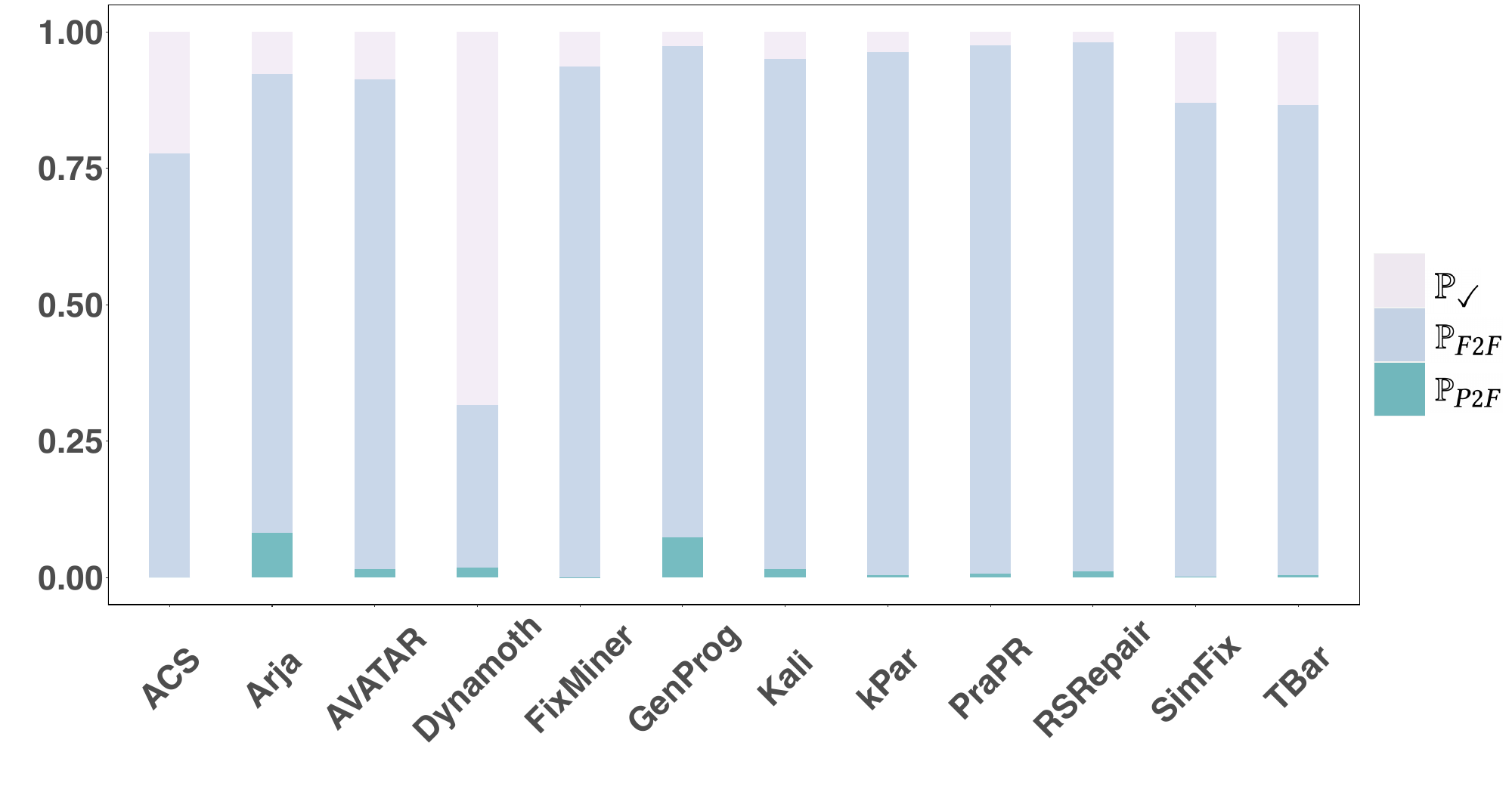}
    \caption{Ratio of patches of different fixing capabilities}
    \label{fig:ratio1}
\end{figure}

Considering the drastic impact of RTS on \plausiblepatch{} and \stopppatch{}, we further check whether an APR system is more susceptible to RTS if it exhibits a higher ratio of \plausiblepatch{} or \stopppatch{}.  Figure~\ref{fig:ratio1} presents the ratio of different patch categories.


First, it is noteworthy that almost all the existing APR systems exhibit a very high ratio (e.g., around 90\%) of \stopfpatch{}, indicating that  \stopppatch{} and \plausiblepatch{} are sparse in the space of compilable patches. 
Liu ~\etal{}~\cite{DBLP:conf/icse/0001WKKB0WKMT20} empirically compared the number of un-compilable patches and inplausible patches generated by different APR systems, to measure their costs wasted in patch validation towards generating a valid patch. 
Their work showed that the state-of-the-art APR techniques can avoid generating un-compilable patches.
However in our work, we make the first attempt to investigate patches of different fixing capabilities (e.g., how many tests can pass on each patch).
Our results suggest that for most existing APR systems, the majority of their inplausible patches have a low capability of fixing originally-failed tests and get immediately terminated in the very beginning stage of validation. 
We also find that the two constraint-based APR systems (e.g., \dynamoth{} and \acs{}) generate a lower ratio of \stopfpatch{}. The potential reason may be that constraint-based APR systems target at conditional expressions, which are more likely to impact the test execution outcomes, i.e., making the originally-failed test pass. In addition, \acs{} designs ranking strategies on fixing ingredients for condition synthesis, which might bring the plausible patch forward, end up the whole validation process earlier, and thus reduce the ratio of \stopfpatch{}.
\Comment{In summary, our results call for more efforts from future work to generate less \stopfpatch{}. Otherwise, unnecessary costs would be spent on compiling and validating these patches. }

Second, there is no apparent correlation between the ratio of \stopppatch{} (or \plausiblepatch{}) and the impact degree of RTS strategies (e.g., with -0.09 Pearson correlation coefficient). For example, although \prapr{} exhibits a quite low ratio of  \stopppatch{} and \plausiblepatch{} (e.g., 0.29\% and 0.93\%), it reduces a larger portion of test executions than other APR systems. 
One straightforward reason is that the number of originally-failed tests is significantly less than the originally-passed tests. For example, each buggy version in Closure has  2.63 failed tests and 7,180.89 passed tests on average.
The number of skipped originally-passed tests is often significantly larger than the number of originally-failed tests, and thus the impact of RTS on originally-passed tests can still dominate the overall efficiency. 
It also explains why even with such a low ratio of \stopppatch{} and \plausiblepatch{} we can still observe a remarkable impact of RTS on overall  efficiency.


\finding{\textit{the majority of compilable patches generated by existing APR systems immediately fail on the originally-failed tests\Comment{, and the constraint-based APR systems generate a higher ratio of patches that can fix originally-failed tests}. Even with such a low ratio of \stopppatch{} and \plausiblepatch{}, the APR efficiency is still sensitive to RTS strategies, since most bugs exhibit substantially more originally-passed tests than originally-failed tests.}}

\begin{table}[htb]
    \centering
   \caption{Reduction on single/multiple edit patches}\label{table:rq2_edit}

   	\begin{adjustbox}{width=0.58\columnwidth}

    \begin{tabular}{|r|| rr| rr| rr|}\hline
    
    \multirow{2}{*}{Subject} & \multicolumn{2}{c|}{\rtsclass{}} & \multicolumn{2}{c|}{\rtsmethod{}} & \multicolumn{2}{c|}{\rtsstate{}} \\  \cline{2-7}
    
    & $\patches_{MC}$ & $\patches_{SC}$ & $\patches_{MM}$ & $\patches_{SM}$  & $\patches_{MS}$ & $\patches_{SS}$ \\ \hline \hline  

Lang&21.83\%&27.36\%&21.83\%&27.86\%&15.25\%&25.75\%\\
Math&26.18\%&28.82\%&17.67\%&30.35\%&20.45\%&29.75\%\\
Time&9.85\%&13.26\%&20.78\%&16.49\%&7.53\%&14.77\%\\
Chart&32.92\%&30.99\%&27.95\%&35.14\%&19.27\%&33.64\%\\
Closure&-&11.30\%&-&19.48\%&55.06\%&17.34\%\\
Mockito&-&9.48\%&-&14.62\%&- &15.62\%\\

 \hline
 
    \end{tabular}
    \end{adjustbox}

\end{table}

\subsubsection{Impact on patches of different fixing scopes} \label{sec:cate2}

Based on the scope of fixed code, all the generated patches can be categorized as single-edit or multiple-edit patches. 
In particular, according to the granularity of edited code elements, we categorize patches as: (1) patches editing single class (denoted as \scpatch{}) v.s. patches editing multiple classes (denoted as \mcpatch{})\Comment{change the symbols into $\mathbb{P}_{C}$ and $\mathbb{P}_{\vec{C}}$?}, (2) patches editing single method (denoted as \smpatch{}) v.s.  patches editing multiple methods (denoted as \mmpatch{}), and (3) patches editing single statement (denoted as \sspatch{}) v.s. patches editing multiple statements (denoted as \mspatch{}).

\begin{figure}[htb]
    \centering
    \includegraphics[scale=0.12]{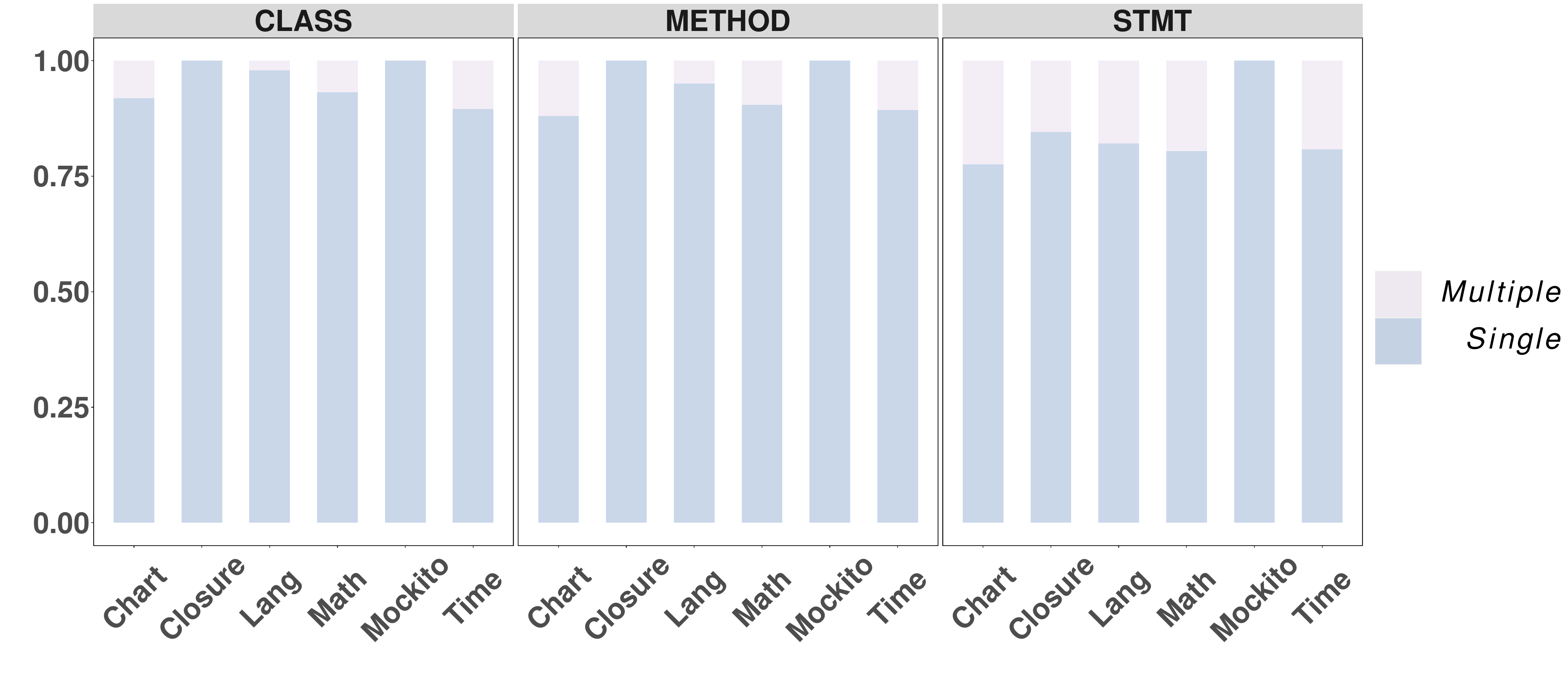}
    \caption{Ratio of patches of different fixing code scopes}
    \label{fig:ratio2}
\end{figure}

Figure~\ref{fig:ratio2} presents the ratio of patch categories of different fixing code scopes.
It shows that the majority of generated patches are single-edit patches,  especially when  edited code elements are at coarser granularities (i.e., class or method).

Table~\ref{table:rq2_edit} presents the reduction achieved by RTS at the corresponding granularity on single-edit and multiple-edit patches. For example, the cell in the column ``\rtsclass{}'' and the row ``Lang'' means that on Lang \rtsclass{} can reduce 21.83\% and 27.36\% of test executions on patches that edit multiple classes  (\mcpatch{}) and on patches that edit single class (\scpatch{}), respectively.
Based on the table, at most cases, RTS can reduce more test executions on single-edit patches than multiple-edit patches. The reason is that multiple-edit patches often involve more code elements and thus are covered by more tests, resulting in a lower ratio of reduction. 
Note there are several counter cases, e.g., on Closure the reduction on multiple-statement patches is much larger than on single-statement patches (55.06\% v.s. 17.34\%). We find that at these cases a considerable ratio of generated multiple-edit patches have stronger fixing capabilities (i.e., belong to \stopppatch{} or \plausiblepatch{}) and thus exhibit a higher reduction.

\finding{\textit{most patches generated by existing APR systems are single-edit patches. RTS can reduce more test executions on single-edit patches than on multiple-edit patches, since usually fewer tests can cover the modified code elements of single-edit patches. Meanwhile, for some special cases, multiple-edit patches can have high probabilities in producing high-quality patches passing more tests, leaving more room for RTS.}}

\subsection{RQ4: Impact with the Full Matrix}~\label{sec:rq4}
In RQ1-RQ3, we investigate the impact of RTS with the partial validation matrix ($\pMatrixpart{}$), i.e., the early-exit mechanism is enabled and the validation for each patch would be terminated after any test failure. In RQ4, we study the impact of RTS with the full matrix ($\pMatrixfull{}$), i.e., the validation for each patch would continue even when there are failed tests. A full validation matrix is common in the scenario that execution results of all tests are required, such as  unified debugging~\cite{DBLP:journals/ac/Lou0ZH19, benton2020effectiveness}. \revised{In this RQ, we further investigate whether there is any difference of RTS impacts in two scenarios (full validation matrices v.s. partial validation matrices).}

Table~\ref{table:rq3_fullmatrix} shows the accumulated number of test executions $NT_{num}(\pMatrixfull{})$ on each APR system. We present the average number of all subjects and buggy versions. It is notable that the full patch validation matrix collection is extremely resource consuming, e.g., from thousands to millions of tests are executed when there is no RTS. 
We can observe a prominent reduction when RTS strategies at finer granularities are integrated in APR systems. For example, on average, \emph{for each buggy version}, \rtsstate{} even reduces millions of test executions for \prapr{}.
In summary, our results show that RTS should definitely be applied in the full-matrix scenario.

\finding{\textit{the full validation matrix is extremely resource-consuming, for which RTS has a more remarkable impact by reducing even up to millions of test executions.}}

\begin{table}[htb]
    \centering
   \caption{Number of test executions with full matrices}\label{table:rq3_fullmatrix}

   	\begin{adjustbox}{width=0.6 \columnwidth}
    \begin{tabular}{|r|rrrr|}\hline

    APR & \norts{} &  \rtsclass{}  & \rtsmethod{}  & \rtsstate{} \\ \hline \hline
    
    PraPR&20,018,923.45& 12,495,932.50& 9,958,460.34& 9,600,813.31\\
SimFix& 70,588.42& 11,408.54& 1,483.51& 1,039.05\\
AVATAR&11,065.76&2,821.98&1,219.36&591.52\\
kPar&9,903.78&2,435.97&1,088.50&509.45\\
TBar&10,256.57&2,856.96&1,237.31&522.12\\
FixMiner&6,441.49&740.07&144.00&32.86\\
Dynamoth&596.65&114.32&17.75&16.86\\
ACS&93.46&5.12&1.44&0.71\\
Arja&1,435,025.26&177,282.34&32,882.52&23,861.27\\
Kali&54,751.70&3,999.24&915.59&755.44\\
GenProg&1,585,441.03&199,088.14&33,824.91&23,840.29\\
RSRepair&668,634.75&48,785.24&7,650.52&5,131.89\\

    \hline
    \end{tabular}
    \end{adjustbox}

\end{table}

\begin{figure}[htb]
    \centering
   \includegraphics[scale=0.12]{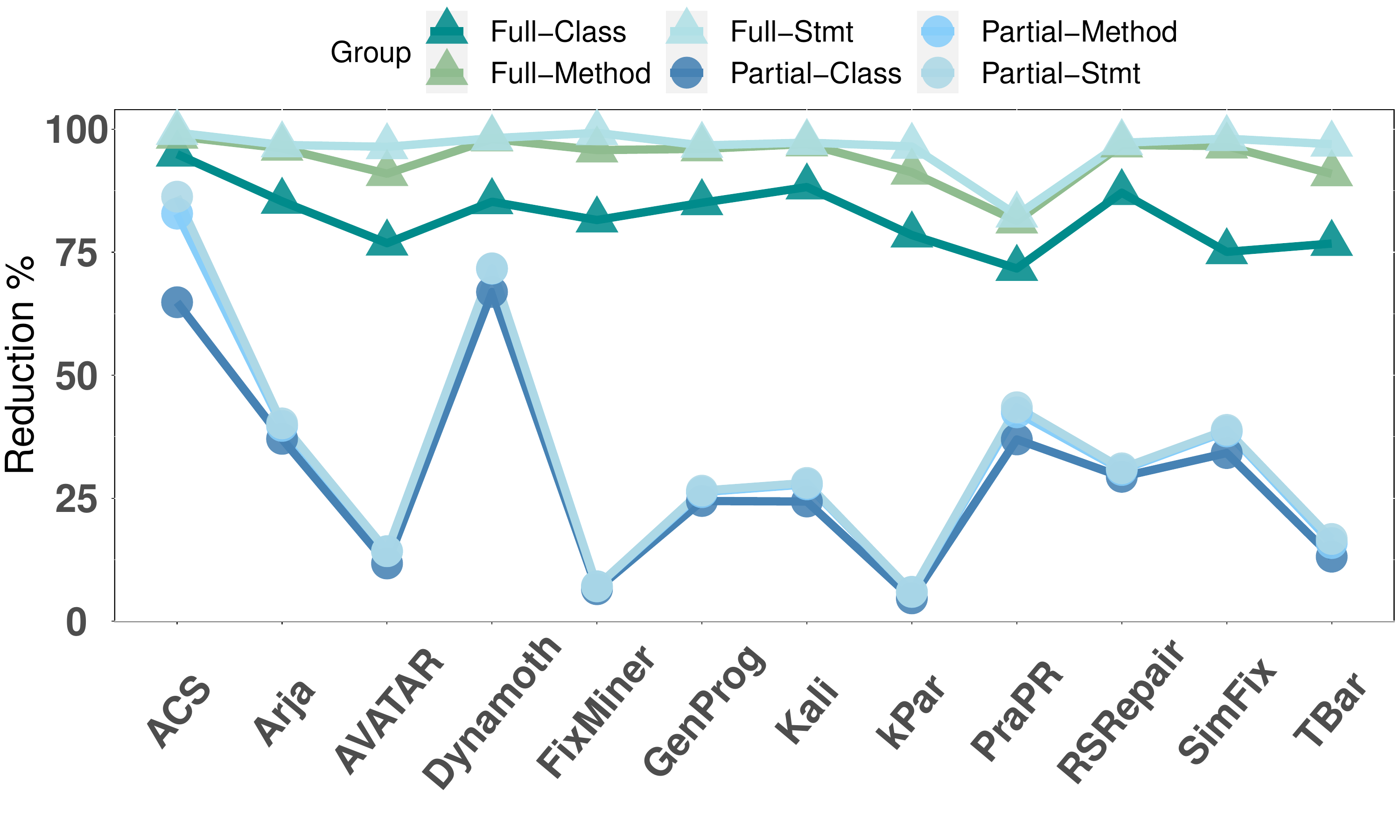}
    \caption{Reduction with full/partial matrices}
    \label{fig:fullpartial}
\end{figure}

We further compare the reduction achieved by RTS between full and partial matrices in Figure~\ref{fig:fullpartial}. 
First, the reduction curves of the full matrix are far beyond the partial matrix. For example, with the full matrix, all the RTS strategies can help all APR systems save more than 70\% of test executions. In particular, \rtsstate{} and \rtsmethod{} can save more than 95\% of  test executions at most cases. 
Second, the difference between partial and full matrices varies among different APR systems. For example, \dynamoth{} achieves a slightly larger reduction while \avatar{} achieves a much larger reduction after they switch from partial matrices into full matrices.
\Comment{Third, the reduction curves are more stable among APR systems with the full matrix than with the partial matrix, indicating that RTS plays an essential role in efficiency optimization especially in the full-matrix scenario. }
Furthermore, consistent with the finding for partial matrices, method- and statement-level RTS perform similarly, and both substantially outperform class-level RTS. Interestingly, their superiority over class-level RTS is even enlarged with full matrices. In fact, the early-exit mechanism in partial matrices prevents more tests from executing, which mitigates the impact from RTS strategies. Therefore, the reduction achieved in the full-matrix scenario represents the upper bound of the impact.

\finding{\textit{RTS can remarkably reduce more test executions with full matrices than with partial matrices. Method- and statement-level RTS still perform similarly, but their superiority over class-level RTS is further enlarged with full matrices.  
\Comment{
the difference between different RTS strategies is also more prominent with the full matrix. The difference of RTS impact among different APR systems is narrowed when switching from partial to full matrices. } \Comment{The difference among APR systems of the impact caused by RTS strategies, is narrowed when switching from partial matrix to full matrix.}}}

\revised{In summary, the impact of RTS in the full matrices differ from the partial matrices in the following two folds. First, the reduction ratio in the full matrices is significantly larger than that in the partial matrices, indicating that RTS is a more necessary efficiency optimization setting for the full-matrix scenario. Second, the gap between the class-level RTS and method-level/statement-level RTS in the full-matrix scenario is substantially larger than that in the partial-matrix scenario, indicating that method-level/statement-level RTS is much more recommended for the full-matrix scenario.}

\begin{table*}[htb]
    \centering
   \caption{Reduction of combining selection and prioritization }\label{table:rq4_combine}

   	\begin{adjustbox}{width=0.98 \columnwidth}

    \begin{tabular}{|r|rrr|rrrr|rrrr|rrrr|}\hline
    
   \multirow{2}{*}{APR} &  \multicolumn{3}{c|}{\norts{}  (\%)} &  \multicolumn{4}{c|}{\rtsclass{} (\%) } &  \multicolumn{4}{c|}{\rtsmethod{} (\%) } &  \multicolumn{4}{c|}{\rtsstate{} (\%) } \\ \cline{2-16}

     & \priotrp{} & \priototal{} & \prioadd{} & \priodefault{} & \priotrp{} & \priototal{} & \prioadd{}  & \priodefault{} & \priotrp{} & \priototal{} & \prioadd{}  & \priodefault{} & \priotrp{} & \priototal{} & \prioadd{} \\  \hline \hline

PraPR&14.53&5.08&\cellcolor{gggray}15.21&36.96&\cellcolor{gggray}41.44&35.44&40.38&42.47&44.55&41.13&\cellcolor{gggray}44.95&43.50&\cellcolor{gggray}44.82&42.91&44.35\\

SimFix &\cellcolor{gggray}20.51&-33.17&14.21&34.19&\cellcolor{gggray}36.76&14.99&36.02&38.58&\cellcolor{gggray}39.25&20.86&38.74&38.90&\cellcolor{gggray}39.33&37.20&39.03\\

AVATAR&0.98&1.44&\cellcolor{gggray}4.31&\cellcolor{gggray}11.75&11.28&11.48&10.47&14.23&14.24&14.24&\cellcolor{gggray}14.31&14.25&14.26&14.25&\cellcolor{gggray}14.30\\
kPar&-1.75&-0.83&\cellcolor{gggray}0.66&\cellcolor{gggray}4.66&3.81&3.93&4.21&5.96&5.98&\cellcolor{gggray}5.99&4.43&5.99&6.00&\cellcolor{gggray}6.01&5.90\\
TBar&-1.08&-1.49&\cellcolor{gggray}2.79&13.13&11.96&11.66&\cellcolor{gggray}13.86&\cellcolor{gggray}15.90&\cellcolor{gggray}15.90&15.89&15.44&\cellcolor{gggray}16.73&\cellcolor{gggray}16.73&16.72&16.14\\
FixMiner&-1.44&-0.62&\cellcolor{gggray}0.08&6.52&\cellcolor{gggray}6.54&6.51&6.07&7.09&\cellcolor{gggray}7.10&\cellcolor{gggray}7.10&6.64&7.11&7.11&7.11&\cellcolor{gggray}8.53\\
Dynamoth&2.07&1.49&\cellcolor{gggray}5.44&66.93&\cellcolor{gggray}67.01&66.95&66.89&71.66&\cellcolor{gggray}71.71&71.67&71.65&71.74&\cellcolor{gggray}71.79&71.74&71.77\\

ACS &\cellcolor{gggray}0.00&-33.16&-4.12&64.84&64.84&64.42&\cellcolor{gggray}65.93&85.82&85.82&85.85&\cellcolor{gggray}85.97&86.30&86.30&\cellcolor{gggray}86.32&\cellcolor{gggray}86.32\\

Arja&\cellcolor{gggray}20.58&5.33&13.30&37.08&39.33&37.71&\cellcolor{gggray}39.71&39.71&40.72&40.01&\cellcolor{gggray}40.92&40.22&\cellcolor{gggray}40.79&40.48&40.62\\
Kali&6.80&6.90&\cellcolor{gggray}10.29&24.34&25.81&25.29&\cellcolor{gggray}26.16&27.83&28.10&27.87&\cellcolor{gggray}28.98&28.12&\cellcolor{gggray}28.17&28.13&28.14\\
GenProg&\cellcolor{gggray}14.03&7.85&11.51&24.45&25.79&24.57&\cellcolor{gggray}25.87&26.24&26.62&26.11&\cellcolor{gggray}27.14&26.55&\cellcolor{gggray}26.66&26.49&26.56\\
RSRepair&\cellcolor{gggray}14.31&9.00&10.95&29.36&30.47&29.71&\cellcolor{gggray}31.08&30.75&31.04&30.79&\cellcolor{gggray}32.02&31.09&\cellcolor{gggray}31.13&31.06&31.11\\

\hline 
    \end{tabular}
    \end{adjustbox}

\end{table*}

\subsection{RQ5: Test Selection+Prioritization}~\label{sec:rq5}
So far we have studied the RTS impact on APR efficiency controlled in a default test execution order (i.e., the lexicographic order).
\Comment{and our results suggest that RTS can largely reduce the number of test executions and improve APR efficiency.} In fact, when the early-exit mechanism is enabled (the default setting for most modern APR systems), the test execution order can also affect the final number of test executions since it decides when the first failure-triggering test would be executed. 
Therefore, we further study the impact of different test prioritization strategies and their joint impact with RTS on APR efficiency:\Comment{. For comparison, we also consider other test prioritization strategies besides RTP techniques as follows.}
 \begin{itemize}
    \item \textbf{Baseline Prioritization.} Tests are scheduled by their lexicographic order, which is also the default setting in previous RQs (denoted as \priodefault{}).
    
    \item \textbf{Patch-history-based Prioritization.} Qi~\etal{}~\cite{DBLP:conf/icsm/QiML13} consider the tests that have failed on more validated patches as more likely to fail on the current patch and schedule them to execute earlier.
    We consider this approach since it is the state-of-the-art test prioritization strategy specifically designed for APR, and we are the first to study its combination with RTS (denoted as \priotrp{}).
    
   \item \textbf{Regression Test Prioritization (RTP).} 
   \Comment{ Test prioritization techniques have also been widely studied in traditional regression testing to reorder regression tests for early detection of regression bugs~\cite{DBLP:journals/tse/LiHH07, DBLP:journals/ac/Lou0ZH19, DBLP:conf/sigsoft/WangNT17}. RTP schedules the execution order of regression test cases to advance the time of failure exposure, which can be naturally adopted to boost patch validation.} 
   Besides RTS, RTP techniques have also been widely studied in traditional regression testing to reorder tests for early detection of regression bugs~\cite{DBLP:journals/tse/LiHH07, DBLP:journals/ac/Lou0ZH19, DBLP:conf/sigsoft/WangNT17}.
   Since each patch can be treated as a new revision in regression testing, we can  naturally apply RTP  for APR. In particular, we consider two most widely studied approaches, i.e., the \emph{total} and \emph{additional} RTP techniques based on statement coverage~\cite{DBLP:conf/icsm/RothermelUCH99}. They share the similar intuition that tests covering more statements or more yet-uncovered statements are more likely to detect regression bugs (denoted as \priototal{} and \prioadd{}). \emph{Note that we are also the first to directly apply RTP for APR.}
\end{itemize}

For each patch, RTS is applied to decide the tests for execution and test prioritization is applied to decide their execution order.
We still follow the common practice of all recent APR systems that all the originally-failed tests are executed before all originally-passed tests, because the former are more likely to fail again on patches. Within the originally-failed tests or the originally-passed tests, we further apply the  test prioritization strategies to schedule their execution order. 

In Table~\ref{table:rq4_combine},  we present the reductions achieved by the combination of different RTS and prioritization strategies when compared to \norts{} with \priodefault{}. The highest reduction within each RTS strategy is highlighted. Based on the table, we have the following observations. 
First, combining RTS with test prioritization can further improve  APR efficiency. However, compared to RTS, test prioritization brings much lower reduction ratios.\Comment{strategies bring smaller benefits in terms of the reduction ratio. In other words, the pure reduction from prioritization is much smaller than from RTS.} For example, when \prapr{} adopts no RTS, the best prioritization (i.e., \prioadd{}) can achieve only 15.21\% reduction, but the worst RTS alone (i.e., \rtsclass{} with \priodefault{}) can achieve 36.96\% reduction. 
This indicates that test selection strategies play a more essential role in efficiency optimization.
Second, the best prioritization strategy varies when combined with different RTS strategies. For most APR systems, surprisingly, the traditional RTP strategy \prioadd{} reduces the most test executions when combined with \norts{}, \rtsclass{}, or \rtsmethod{}. We further perform Wilcoxon-Signed-Rank Test at the significance level of 0.05 and find the differences are all statically significant (i.e., p-values $<$ 0.05).
To the best of our knowledge, this is the first study demonstrating that in the APR scenario, the traditional RTP technique even outperforms state-of-the-art APR-specific test prioritization technique at the most cases (e.g., without RTS and with coarse-grained RTS).
One potential reason may be that the generated patches share little commonality and it is challenging for \priotrp{} to infer the results of un-executed patches from historical patch executions. In contrast, \prioadd{} executes tests with diverse statement coverage as early as possible and thus can advance buggy patch detection.
Furthermore,  it is notable that \prioadd{} becomes less effective than \priotrp{} when combined with \rtsstate{}. 
Intuitively, the tests selected by coarser-grained coverage criteria (e.g., class-level RTS) tend to exhibit a more diverse distribution of statement coverage than the tests selected by the statement-level RTS. Therefore, when combined with \rtsstate{}, \priotrp{} has a stronger capability in distinguishing tests based on their historical execution results.
Third, we can observe that \prioadd{} outperforms \priototal{} in terms of the repair efficiency, which is consistent with the previous findings in traditional regression testing\Comment{, i.e., the additional strategy usually outperforms the total strategy}~\cite{DBLP:conf/icse/LuLCZHZ016, DBLP:conf/icsm/RothermelUCH99}.

\finding{\textit{ 
this study demonstrates for the first time that test selection outperforms test prioritization for APR, and their combination could further improve APR efficiency. Also, surprisingly, traditional RTP strategy \prioadd{} outperforms state-of-the-art APR-specific test prioritization \priotrp{} for most cases.}}

\section{Discussion\Comment{on Time Cost}}~\label{sec:dis}

\begin{figure}[htb]
    \centering
   \includegraphics[width=6.5cm, height= 3.2 cm]{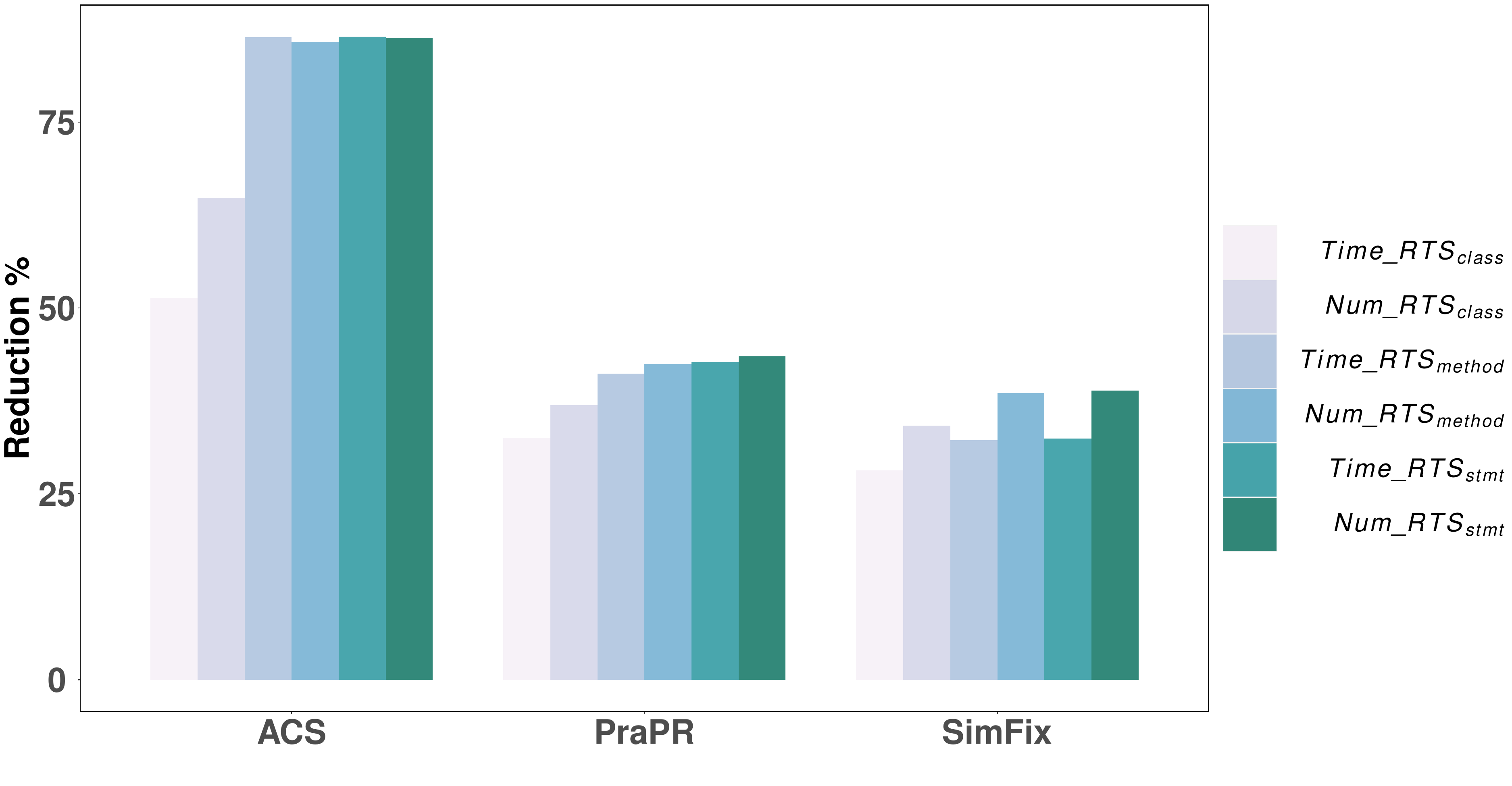}
    \caption{Reduction in test time vs.  \# of test executions}
    \label{fig:cmp_time}
\end{figure}

\parabf{Time Costs.} 
\revised{Figure~\ref{fig:cmp_time} presents the reduction of both the time and the number of test executions achieved by different RTS strategies on three representative APR systems (i.e., \prapr{}, \simfix{}, and \acs{}). In particular, we choose \prapr{}, \simfix{}, and \acs{} as they are the latest APR techniques in their belonging categories (i.e., template-based, heuristic-based, and constraint-based APR). Based on the figure, we could have the following observation. In particular, the trends are actually consistent between the time costs and the number of test executions. For example, on \acs{}, the gap between class-level RTS strategy and other RTS strategies is much larger in terms of both time costs and number of test execution. In fact, such an observation is as expected, as the number of tests in a test suite is often very large and the reduction achieved by RTS is significant. Note that here we already include the RTS overheads into time costs, but it is rather lightweight compared to the patch execution time. For example, for PraPR with the largest project Closure, RTS takes 13 seconds for all patches while the patch execution without RTS takes about 4 hours. Hence, the number of reduced test executions is so large that the diversity in the execution time of different tests would have little effect on the results. In other words, although different tests may have different execution time, the reduction achieved by RTS strategies is so significant that it exhibits similar trends between the number of test executions and test execution time.}
\revised{In summary, the time costs and number of test executions can be alternative when measuring the efficiency of APR systems. In particular, time costs can demonstrate the efficiency in a more straightforward way while the  nondeterminism is supposed to be mitigated by multiple executions; the number of executions can demonstrate the efficiency in a more stable way. Our results encourage the community to consider test executions as a status quo metric in the future work.}


\Comment{2/40 hours\lingming{the number is wrong, as they are with statement rts...}.}
\Comment{More specifically, RTS costs mainly consist of coverage collection, diff computation, and the actual RTS analysis (simply matching coverage against diffs). Since for APR detailed coverage is already collected by the fault-localization component, and diffs are also recorded during patch generation, the actual RTS analysis is extremely lightweight.}

\begin{figure}[htbp]
\centering
\subfigure{
\includegraphics[scale=0.09]{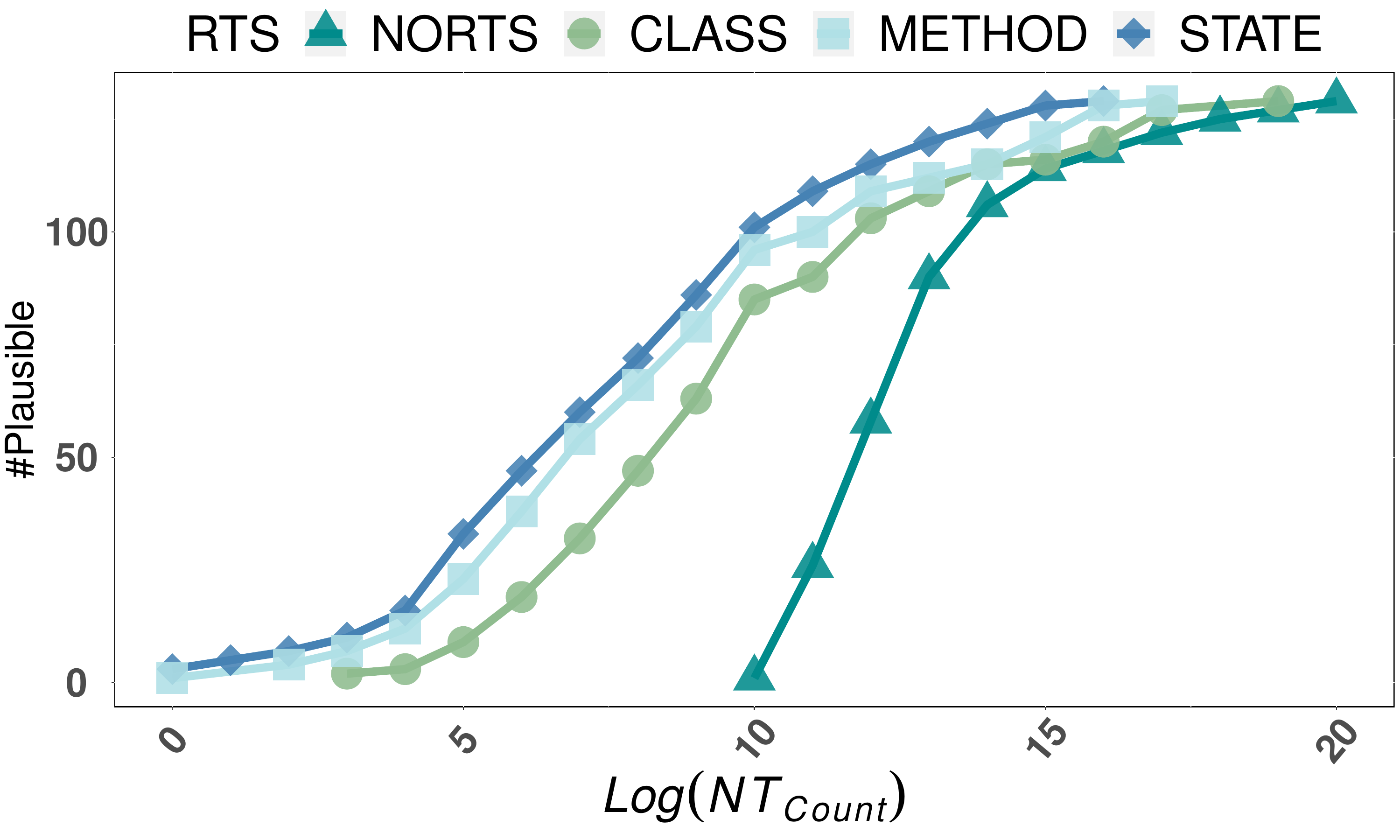}~\label{fig:plausible1}
}
\subfigure{
\includegraphics[scale=0.09]{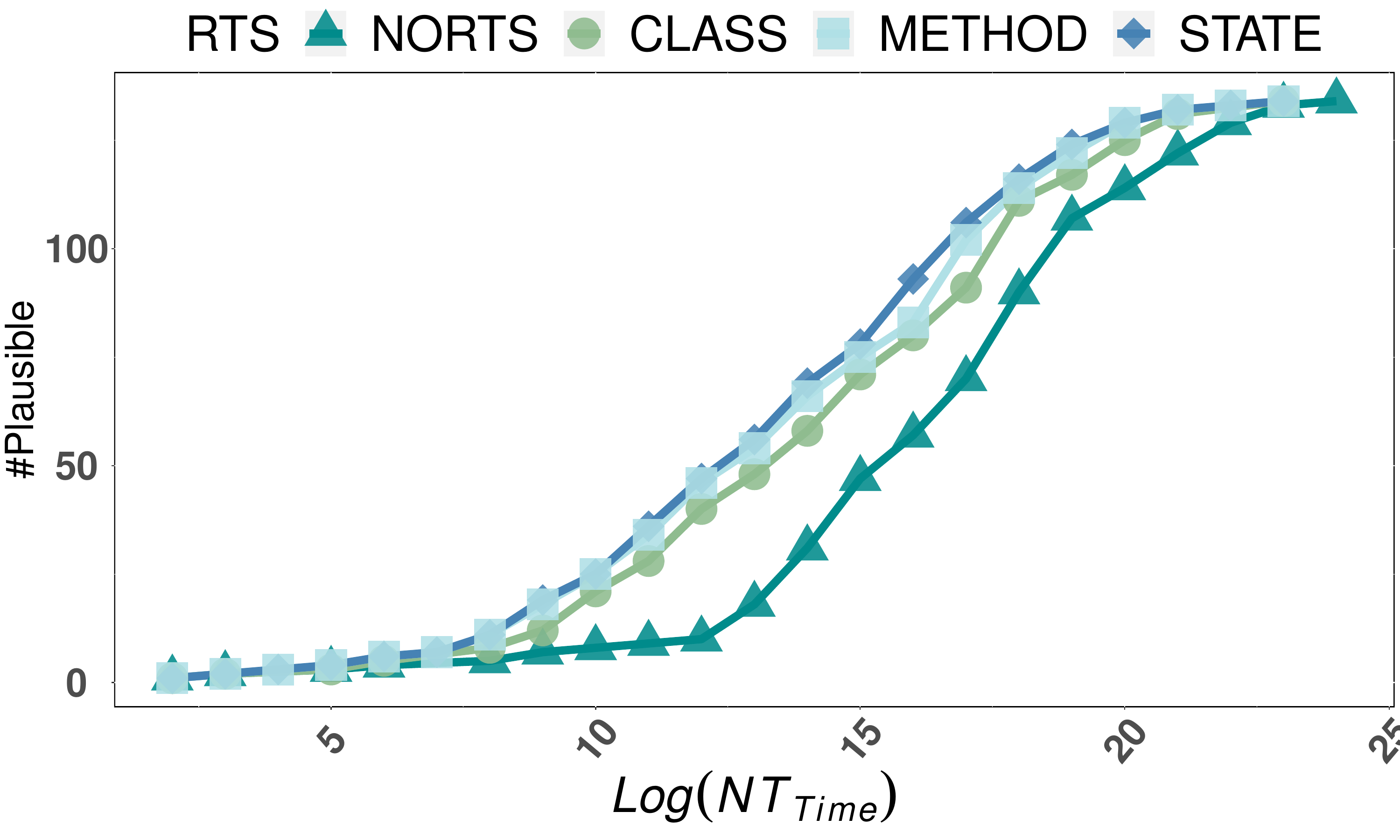}~\label{fig:plausible2}
}
\caption{Plausible Patches with different RTS strategies}~\label{fig:plausible}
\end{figure}

\parabf{Repair Effectiveness.} We further discuss the impact of different RTS strategies on repair effectiveness. Figure~\ref{fig:plausible} shows the number of plausible patches (the y-axis) found by different number/time of test executions (the x-axis). For space limits, we present \prapr{} results here, and other results are in our website. The figure indicates a consistent observation with our previous findings: RTS improves APR efficiency, and thus it helps find more plausible patches within the same budgets.

\section{Related work}\label{sec:related}
Since Section~\ref{sec:motivation} presents background about RTS and APR, here we focus on other regression testing techniques and the closely related work for improving APR efficiency.

\parabf{Regression Testing.} Besides RTS, researchers also propose other two categories of regression testing techniques, i.e., Regression Test Prioritization (RTP)~\cite{DBLP:journals/tse/LiHH07, DBLP:journals/ac/Lou0ZH19, DBLP:conf/icsm/QiML13} and Test-Suite Reduction (TSR)~\cite{DBLP:journals/stvr/YooH12}. With the common goal of accelerating regression testing, RTP  reorders test executions for earlier fault detection, while TSR removes \emph{redundant} tests permanently according to certain testing requirements. Since it is often challenging to identify \emph{redundant} tests, TSR can incur fault detection loss and is not as widely adopted (compared to RTS and RTP). 
This study excludes TSR because it may produce incorrect patch validation results and reduce the APR accuracy. 

\parabf{APR Efficiency.} APR is expensive due to the large number of generated patches and non-trivial costs in patch validation. In addition to many APR approaches that aim to reduce the number of generated patches (mentioned in Section~\ref{sec:apr}), researchers propose to reduce the number of test executions for each patch\Comment{ by test case prioritization, test suite reduction, and regression test selection}. 
For example, Qi~\etal{}~\cite{DBLP:conf/icsm/QiML13} \Comment{ and Venugopal~\etal{}~\cite{DBLP:conf/icuimc/JangPL19, app10051593}} utilize patch execution history to prioritize tests.
\Comment{Fast~\etal{}~\cite{DBLP:conf/gecco/FastGFW10} performed the random and time-aware test suite reduction on the originally-passed tests to accelerate patch validation.}
Mehne~\etal{}~\cite{DBLP:conf/icst/MehneYPSGK18} reduce test executions based on statement coverage (i.e., statement-level RTS); similarly, as suggested by Table~\ref{table:revisit}, several existing APR systems also leverage RTS to accelerate patch validation, e.g., the ARJA family~\cite{DBLP:journals/tse/YuanB20/arja} adopts statement-level RTS and \capgen{}~\cite{DBLP:conf/icse/WenCWHC18} adopts class-level RTS.
\revised{In addition, Mechtaev et al.~\cite{DBLP:journals/tosem/MechtaevGTR18} and Just et al.~\cite{DBLP:conf/issta/JustEF14} skip redundant test executions based on program-equivalence, i.e., the test executions of two test-equivalent patches (e.g., exhibiting indistinguishable test results) can be essentially reduced. } 
Lou et al.~\cite{DBLP:journals/tse/BentonLLZ22, DBLP:conf/kbse/BentonLLZ20, DBLP:conf/issta/LouGLZZHZ20} unify the fault localization and program repair to narrow down the search space of the buggy location, so as to further facilitate the debugging process.

\Comment{Different from the APR context, it is interesting that a recent study also investigated RTS impact on non-functional genetic improvement~\cite{DBLP:conf/icse/GuizzoPSH21}. The previous work suggested a promising efficiency improvement achieved by test execution optimization strategies, while our work performs the first study to extensively investigate the impact of representative regression testing techniques on a wide range of state-of-the-art APR systems (i.e., on 12 APR systems and over 2M patches).} However, the community still lacks a comprehensive understanding on the benefits of RTS for APR, and our work conducts the first study to systematically investigate the impact of representative RTS techniques on a wide range of state-of-the-art APR systems.

Besides reducing the number of test executions, researchers have also looked into speeding up the time for each test/patch execution during APR. For example, JAID~\cite{DBLP:conf/kbse/Chen0F17/jaid} and SketchFix~\cite{DBLP:conf/icse/HuaZWK18/sketchfix} transformed the buggy programs into meta-programs or sketches\Comment{ (i.e., abstract program states or partial program with holes)} to accelerate patch validation\Comment{by pruning redundant test executions}. 
Ghanbari~\etal{}\cite{DBLP:conf/issta/GhanbariBZ19}\Comment{ highlighted the non-negligible costs in patch compilation and} proposed a bytecode-level APR approach which requires no patch compilation and system reloading. More recently, Chen \etal{} ~\cite{chen2020fast} leveraged on-the-fly patch validation to save patch loading and execution time to speed up all existing source-code-level APR techniques. Such techniques are orthogonal to test execution reduction studied in this work, and they can be combined to further reduce APR cost.

\Comment{In addition to cost reduction techniques, researchers have also paid attention to the efficiency measurement in APR systems. 
Most of existing work assessed repair efficiency based on the CPU time spent during repair process, which has been suggested as unstable and biased in previous work~\cite{DBLP:conf/icse/0001WKKB0WKMT20}. 
Besides time costs, Ghanbari \etal{} ~\cite{DBLP:conf/issta/GhanbariBZ19} reported the number of patches generated by the APR system PraPR. 
Liu \etal{}~\cite{DBLP:conf/icse/0001WKKB0WKMT20} further performed an extensive study to measure the efficiency of existing 16 APR systems based on the number of generated patches. 
Compared to previous work, our work makes the first attempt to study the APR efficiency based on the number of test executions. In addition, we also address potential threats from the inconsistent RTS strategies in existing APR efficiency work. Our results reveal that the number of patches widely used for measuring APR efficiency can incur skewed conclusions, and inconsistent RTS configurations can further skew the conclusion. 

}

\section{Conclusion and Future Work}\label{sec:conclusion}
This work points out an important test execution optimization (RTS) largely neglected and inconsistently configured by existing APR systems. We perform the first extensive study of different RTS techniques for 12 state-of-the-art APR systems on over 2M patches. Our findings include: the number of patches widely used for measuring APR efficiency and the inconsistent RTS configurations can both incur skewed conclusions; all studied RTS techniques substantially improve APR efficiency, while method/statement-level RTS significantly outperform class-level RTS. We also present the RTS impact on different patches and its combination with test prioritization.

\revised{In the future, we plan to extend this work with more  APR systems and more test selection strategies. In particular, this work currently focuses on traditional APR systems. While given the recent advance in large language models (LLMs), investigating the test execution efficiency in LLM-based APR systems~\cite{DBLP:conf/sigsoft/0003X023, DBLP:conf/icse/XiaWZ23,DBLP:conf/sigsoft/XiaZ22} is an essential problem. In addition, this work mainly focuses on the RTS strategies in patch validation, while it remains many open problems on more customized test selection strategies for patch validation, such as leveraging static dependencies or run-time dependencies for test selection.
In addition, extending this study to more diverse projects and different programming languages can also be interesting future work.}

\Comment{
In this work, we first pointed out an important test execution optimization (RTS) largely neglected and inconsistently configured by existing APR systems. We then performed the first study on the impact of RTS on APR efficiency. More specifically, we conducted our study for three representative RTS strategies on 12 state-of-the-art APR systems and 395 real bugs from the widely-used Defects4J benchmark. Our study has provided various practical guidelines for future APR work, including: (1) the number of patches widely used for measuring APR efficiency can incur skewed conclusions, and the use of inconsistent RTS configurations can further skew the conclusion; (2) all studied RTS techniques can substantially improve APR efficiency and should be considered in future APR work; (3) RTS at finer granularities (such as the method and statement levels) substantially outperforms class-level RTS, and are more recommended; (4) RTS techniques can substantially outperform state-of-the-art test prioritization techniques for APR, and they can be combined to further improve APR efficiency; and (5) traditional regression test prioritization widely studied in regression testing performs even better than APR-specific test prioritization when combined with most RTS techniques. Furthermore, we have also investigated the impact of different patch categories and patch validation strategies on our findings. 
}

\balance
\bibliographystyle{ACM-Reference-Format}
\bibliography{full, pc}

\end{document}